\documentstyle[preprint,aps]{revtex}
\begin{document}
\draft

\title
{The Lienard-Wiechert Potential of Charged Scalar Particles and their Relation 
to Scalar Electrodynamics in the Rest-Frame Instant Form.}

\author{David Alba}

\address
{Dipartimento di Fisica\\
Universita' di Firenze\\
L.go E.Fermi 2 (Arcetri)\\
50125 Firenze, Italy}

\author{and}

\author{Luca Lusanna}

\address
{Sezione INFN di Firenze\\
L.go E.Fermi 2 (Arcetri)\\
50125 Firenze, Italy\\
E-mail LUSANNA@FI.INFN.IT}

\maketitle

\begin{abstract}

After a summary of a recently proposed new type of instant form of dynamics
(the Wigner-covariant rest-frame instant form), the reduced Hamilton equations
in the covariant rest-frame Coulomb gauge for the isolated system of N scalar
particles with pseudoclassical Grassmann-valued electric charges plus the
electromagnetic field are studied. The Lienard-Wiechert potentials of the
particles are evaluated and it is shown how the causality problems of the
Abraham-Lorentz-Dirac equation are solved at the pseudoclassical level. Then,
the covariant rest-frame description of scalar electrodynamics is given. 
Applying to it the Feshbach-Villars formalism, the connection with the particle 
plus electromagnetic field system is found.

\vskip 1truecm
\noindent April 1997
\vskip 1truecm
\noindent This work has been partially supported by the network ``Constrained 
Dynamical Systems" of the E.U. Programme ``Human Capital and Mobility".

\end{abstract}
\vfill\eject

\section
{Introduction}

In a recent paper \cite{lus1} a new type of instant form of dynamics 
\cite{dira2} was introduced with special Wigner-covariance properties and was
named ``covariant rest-frame instant form". The twofold motivations which led 
to its discovery were the problem of understanding the role of relative times 
in the description of N relativistic scalar particles [whose phase space
coordinates are $x^{\mu}_i(\tau )$, $p^{\mu}_i(\tau )$, i=1,..,N] with first
class constraints $p^2_i-m^2_i\approx 0$ [see Ref.\cite{lus1} and the references
quoted there for the solution of this problem] and the need of a formulation
already adapted to the transition from special to general relativity.

The N particle system (starting with N free particle for the sake of simplicity)
was reformulated on spacelike hypersurfaces $\Sigma_{\tau}$, all diffeomorphic
to a given one $\Sigma$, foliating Minkowski spacetime ($\tau$ is a scalar
parameter labelling the leaves of the foliation) following Dirac's approach to
parametrized field theory \cite{dira1}, subsequently extended to curved 
spacetimes by Kuchar \cite{kuchar}. In this way one adds an infinite number of
configuration variables, the coordinates $z^{\mu}(\tau ,\vec \sigma )$ of the
points of $\Sigma_{\tau}$ [$\vec \sigma$ are Lorentz-scalar curvilinear
coordinates on the abstract $\Sigma$ embedded in Minkowski spacetime as $\Sigma
_{\tau}$]. The position in $\Sigma_{\tau}$ of a particle travelling on a
timelike worldline $\gamma_i$ is determined only by the 3 Lorentz-scalar 
numbers $\vec \sigma ={\vec \eta}_i(\tau )$ determining the intersection of
$\gamma_i$ with $\Sigma_{\tau}$: $\, x^{\mu}_i(\tau )=z^{\mu}(\tau ,{\vec \eta}
_i(\tau ))$. But this means that the constraints $p^2_i-m^2_i\approx 0$ have
been solved in this description, so that $p^o_i\approx \pm \sqrt{m^2_i+{\vec p}
_i^2}$ and all the particle time variables $x^o_i(\tau )$ have been replaced by
the $\tau$-value identifying $\Sigma_{\tau}$ [namely all relative times are put
equal to zero from the beginning in a covariant way]. Therefore, in this
1-time description every particle has a well defined sign $\eta_i=\pm 1$ of the
energy and we cannot describe simultaneously both times like in the N-times
description.

The standard Lagrangian of N free paerticles was rewritten in terms of ${\vec 
\eta}_i(\tau )$, ${\dot {\vec \eta}}_i(\tau )={d\over {d\tau}} {\vec \eta}
_i(\tau )$ and of the metric induced on $\Sigma_{\tau}$ by the Minkowski
metric $\eta_{\mu\nu}$, namely of $g_{\check A\check B}(\tau ,\vec \sigma )=
z^{\mu}_{\check A}(\tau ,\vec \sigma )\eta_{\mu\nu}z^{\nu}_{\check B}(\tau 
,\vec \sigma )$, where $z^{\mu}_{\check A}(\tau ,\vec \sigma )={{\partial}\over
{\partial \sigma^{\check a}}}z^{\mu}(\tau ,\vec \sigma )$ [$\, \sigma^{\check 
A}=(\sigma^{\check o}=\tau ; \vec \sigma =\{ \sigma^{\check r}\} )$]. It turns 
out that there are 4 first class constraints determining the momentum $\rho
_{\mu}(\tau ,\vec \sigma )$, conjugate to $z^{\mu}(\tau ,\vec \sigma )$, in
terms of the momenta ${\vec \kappa}_i(\tau )$ conjugate to the ${\vec \eta}
_i(\tau )$'s and of the vectors normal and tangent to $\Sigma_{\tau}$ [all
functions of the $z^{\mu}_{\check r}(\tau ,\vec \sigma )$]: the component of
$\rho_{\mu}(\tau ,\vec \sigma )$ along the normal is the particle
energy-density on $\Sigma_{\tau}$, while the components tangent to $\Sigma
_{\tau}$ are the components of the particle momentum density on $\Sigma_{\tau}$.
These 4 first class constraints say that the description of the N particles is
independent from the choice of the foliation. Therefore, in special relativity
we can get a simpler description by restricting ourselves to foliations whose
leaves are spacelike hyperplanes of Minkowski spacetime. Finally, if we select 
all particle configurations whose total 4-momentum is timelike (they are dense 
in the space of all possible configurations), we can restrict ourselves to the 
special foliation whose hyperplanes are orthogonal to the total 4-momentum:
these hyperplanes, named Wigner hyperplanes $\Sigma_{w\, \tau}$, are
intrinsically determined by the physical system itself (in this case N free
scalar particles). It is shown in Ref.\cite{lus1}, that at this stage all
the degrees of freedom $z^{\mu}(\tau ,\vec \sigma )$, $\rho_{\mu}(\tau ,\vec 
\sigma )$, have disappeared except for the canonical coordinates ${\tilde x}
^{\mu}_s(\tau )$, $p^{\mu}_s$, of a point. While $p^{\mu}_s$ is a timelike
4-vector (playing the role of the total 4-momentum) orthogonal to $\Sigma
_{w\, \tau}$ with $p^2_s\approx squared\, invariant\, mass\, of\, the\, system$
due to the constraints, ${\tilde x}^{\mu}_s(\tau )$ is not a 4-vector. It 
describes the canonical relativistic center of mass of the system: it is the
classical analogue of the Newton-Wigner position operator and, like it, it is 
covariant only under the little group of timelike Poincar\'e orbits (the 
Euclidean group).

As shown in Ref.\cite{lus1}, the restriction to Wigner hyperplanes forces the
Lorentz-scalar 3-vectors ${\vec \eta}_i(\tau )$, ${\vec \kappa}_i(\tau )$, to
become Wigner spin-1 3-vectors [they transform under induced Wigner rotations
when one rotates the Wigner hyperplanes with Lorentz boosts in Minkowski
spacetime], since in the gauge-fixing procedure use is made of the Wigner
standard boost for timelike Poincar\'e orbits. Therefore, tensors on the
Wigner hyperplane have Wigner covariance and the Wigner hyperplanes are
intrisically Euclidean: an 1-time Wigner-covariant relativistic statistical
mechanics can be developed on them as shown in Ref.\cite{lus1}. 

Only 4 first class constraints remain on Wigner hyperplanes: i) one determines 
$\sqrt{p^2_s}$ in terms of the particle-system invariant mass; ii) the other 3 
are  ${\vec p}_W=\sum_{i=1}^N {\vec \kappa}_i(\tau )\approx 0$ [$\, {\vec p}_W$
must not be confused with the space part of $p^{\mu}_s$, which is arbitrary,
being connected with the chosen frame of reference of Minkowski spacetime in
which Wigner hyperplanes are embedded], saying that the Wigner hyperplanes are
the intrinsic rest frames after the separation of the center of mass motion
in Minkowski spacetime. Since the Lorentz-scalar Minkowski-rest-frame time
$T_s=p_s\cdot {\tilde x}_s/\sqrt{p^2_s}$ is the variable conjugate to $\sqrt{
p^2_s}$, one can add the gauge-fixing $T_s-\tau \approx 0$ and obtain a
description of the evolution in $T_s$ of the system by using the invariant
mass as Hamiltonian [the other 6 degrees of freedom in ${\tilde x}^{\mu}_s$, 
$p^{\mu}_s$, are the 6 canonical coordinates of the free noncovariant center of
mass]. If one adds the 3 gauge-fixings $\sum_{i=1}^N{\vec \eta}_i(\tau )\approx
0$, one can reduce the 6N degrees of freedom ${\vec \eta}_i(\tau )$, ${\vec 
\kappa}_i(\tau )$, to 3(N-1) pairs of relative variables [for general systems
usually one does not know the form of the needed 3 gauge-fixings to be added].
In this way one gets the rest-frame instant form of dynamics on the Wigner
hyperplanes and has the relativistic generalization of the Newtonian separation
of the center-of-mass motion in phase space.

On the Wigner hypersurface one can introduce any kind of instantaneous 
action-at-a-distance interactions (without the complications of the N-times
formalism) and to treat the problem of cluster decomposition in Newtonian
terms.

Then in Ref.\cite{lus1} the isolated system of N scalar particles with
Grassmann-valued electric charges plus the electromagnetic field was studied in
this way till the level of Wigner hyperplanes. One also found the Dirac
observables with respect to the gauge transformations of the whole system and
obtained the Wigner-covariant rest-frame version of the Coulomb gauge. In
particular, this allows to extract from field theory the interparticle
instantaneous Coulomb potential (which appears in the invariant mass as an
additive term to the particle relativistic kinetic energies) and to regularize
the Coulomb self-interactions due to the Grassmann character of the electric
charges $Q_i$, which implies $Q^2_i=0$. However, the Hamilton equations of 
motion and their implications in the rest-frame instant form were not given. 
Moreover, even if there were some comments about the relation of charged
particles with the Feshbach-Villars \cite{fv} formalism for charged Klein-
Gordon fields with external electromagnetic fields, no clear connection was
established.

In this paper we shall study the Hamilton equations on the Wigner hyperplane of
the isolated system of N charged scalar particles plus the electromagnetic field
in the pseudoclassical case of Grassmann-valued electric charges of the 
particles. We shall find therest-frame formulation of the Lienard-Wiechert
potentials and the pseudoclassical regularization of the causality problems
of the Abraham-Lorentz-Dirac equation. Each particle has its Lienard-Wiechert
potential, but it does not directly produce radiation because $Q^2_i=0$;
however, one gets a Larmor formula for the radiated energy containing terms
$Q_iQ_j$, $i\not= j$, from the interference of the various Lienard-Wiechert
potentials in wave zone and this is macroscopically satisfactory.

Then, we reformulate scalar electrodynamics, namely the isolated system of a
complex Klein-Gordon field coupled to the electromagnetic field, on
spacelike hypersurfaces and we obtain its rest-frame description on Wigner
hyperplanes. Following Ref.\cite{lv1}, we give the Dirac observables with
respect to the gauge transformations of the theory. Then, we evaluate the 
rest-frame reduced Hamilton equations and we apply to them the Feshbach-Villars
formalism. Finally, we show that one can recover the previous constraints of
chargede particles plus the electromagnetic field from this treatment of
scalar electrodynamics, if one makes a strong eikonal approximation which
eliminates the mixing of positive and negative energy solutions of the Klein-
Gordon theory, which is induced by effects that, in second quantization, are
interpreted as pair production from vacuum polarization.

In Section II we remind some results of Ref.\cite{lus1}. In Section III we
evaluate the reduced Hamilton equations in the rest-frame instant form. In
Section IV we find the Lienard-Wiechert potentials of the particles and we
study their equations of motion, showing which is the pseudoclassical way out 
from the causality problems of the Abraham-Lorentz-Dirac equation. In Section 
V we study scalar electrodynamics on spacelike hypersurfaces and we find its 
rest-frame formulation; then we recast it in the Feshbach-Villars formalism
and look for connections with the previous theory.
Some final comments are put in the Conclusions.

\vfill\eject

\section{Preliminaries}

In this Section we will introduce the background material from Ref.\cite{lus1} 
needed in the description of physical systems on spacelike hypersurfaces, 
integrating it with the definitions needed to describe the isolated system of N
scalar particles with pseudoclassical Grassmann-valued electric charges plus the
electromagnetic field\cite{lus1}.

Let $\lbrace \Sigma_{\tau}\rbrace$ be a one-parameter family of spacelike
hypersurfaces foliating Minkowski spacetime $M^4$ and giving a 3+1 decomposition
of it. At fixed $\tau$, let 
$z^{\mu}(\tau ,\vec \sigma )$ be the coordinates of the points on $\Sigma
_{\tau }$ in $M^4$, $\lbrace \vec \sigma \rbrace$ a system of coordinates on
$\Sigma_{\tau}$. If $\sigma^{\check A}=(\sigma^{\tau}=\tau ;\vec \sigma 
=\lbrace \sigma^{\check r}\rbrace)$ [the notation ${\check A}=(\tau ,
{\check r})$ with ${\check r}=1,2,3$ will be used; note that ${\check A}=
\tau$ and ${\check A}={\check r}=1,2,3$ are Lorentz-scalar indices] and 
$\partial_{\check A}=\partial /\partial \sigma^{\check A}$, 
one can define the vierbeins

\begin{equation}
z^{\mu}_{\check A}(\tau ,\vec \sigma )=\partial_{\check A}z^{\mu}(\tau ,\vec 
\sigma ),\quad\quad
\partial_{\check B}z^{\mu}_{\check A}-\partial_{\check A}z^{\mu}_{\check B}=0,
\label {a1}
\end{equation}

\noindent so that the metric on $\Sigma_{\tau}$ is

\begin{eqnarray}
&&g_{{\check A}{\check B}}(\tau ,\vec \sigma )=z^{\mu}_{\check A}(\tau ,\vec 
\sigma )\eta_{\mu\nu}z^{\nu}_{\check B}(\tau ,\vec \sigma ),\quad\quad 
g_{\tau\tau}(\tau ,\vec \sigma ) > 0,\nonumber \\
&&g(\tau ,\vec \sigma )=-det\, ||\, g_{{\check A}{\check B}}(\tau ,\vec 
\sigma )\, || ={(det\, ||\, z^{\mu}_{\check A}(\tau ,\vec \sigma )\, ||)}^2,
\nonumber \\
&&\gamma (\tau ,\vec \sigma )=-det\, ||\, g_{{\check r}{\check s}}(\tau ,\vec 
\sigma )\, ||.
\label{a2}
\end{eqnarray}

If $\gamma^{{\check r}{\check s}}(\tau ,\vec \sigma )$ is the inverse of the 
3-metric $g_{{\check r}{\check s}}(\tau ,\vec \sigma )$ [$\gamma^{{\check r}
{\check u}}(\tau ,\vec \sigma )g_{{\check u}{\check s}}(\tau ,\vec 
\sigma )=\delta^{\check r}_{\check s}$], the inverse $g^{{\check A}{\check B}}
(\tau ,\vec \sigma )$ of $g_{{\check A}{\check B}}(\tau ,\vec \sigma )$ 
[$g^{{\check A}{\check C}}(\tau ,\vec \sigma )g_{{\check c}{\check b}}(\tau ,
\vec \sigma )=\delta^{\check A}_{\check B}$] is given by

\begin{eqnarray}
&&g^{\tau\tau}(\tau ,\vec \sigma )={{\gamma (\tau ,\vec \sigma )}\over
{g(\tau ,\vec \sigma )}},\nonumber \\
&&g^{\tau {\check r}}(\tau ,\vec \sigma )=-[{{\gamma}\over g} g_{\tau {\check 
u}}\gamma^{{\check u}{\check r}}](\tau ,\vec \sigma ),\nonumber \\
&&g^{{\check r}{\check s}}(\tau ,\vec \sigma )=\gamma^{{\check r}{\check s}}
(\tau ,\vec \sigma )+[{{\gamma}\over g}g_{\tau {\check u}}g_{\tau {\check v}}
\gamma^{{\check u}{\check r}}\gamma^{{\check v}{\check s}}](\tau ,\vec \sigma ),
\label{a3}
\end{eqnarray}

\noindent so that $1=g^{\tau {\check C}}(\tau ,\vec \sigma )g_{{\check C}\tau}
(\tau ,\vec \sigma )$ is equivalent to

\begin{equation}
{{g(\tau ,\vec \sigma )}\over {\gamma (\tau ,\vec \sigma )}}=g_{\tau\tau}
(\tau ,\vec \sigma )-\gamma^{{\check r}{\check s}}(\tau ,\vec \sigma )
g_{\tau {\check r}}(\tau ,\vec \sigma )g_{\tau {\check s}}(\tau ,\vec \sigma ).
\label{a4}
\end{equation}

We have

\begin{equation}
z^{\mu}_{\tau}(\tau ,\vec \sigma )=(\sqrt{ {g\over {\gamma}} }l^{\mu}+
g_{\tau {\check r}}\gamma^{{\check r}{\check s}}z^{\mu}_{\check s})(\tau ,
\vec \sigma ),
\label{a5}
\end{equation}

\noindent and

\begin{eqnarray}
\eta^{\mu\nu}&=&z^{\mu}_{\check A}(\tau ,\vec \sigma )g^{{\check A}{\check B}}
(\tau ,\vec \sigma )z^{\nu}_{\check B}(\tau ,\vec \sigma )=\nonumber \\
&=&(l^{\mu}l^{\nu}+z^{\mu}_{\check r}\gamma^{{\check r}{\check s}}
z^{\nu}_{\check s})(\tau ,\vec \sigma ),
\label{a6}
\end{eqnarray}

\noindent where

\begin{eqnarray}
l^{\mu}(\tau ,\vec \sigma )&=&({1\over {\sqrt{\gamma}} }\epsilon^{\mu}{}_{\alpha
\beta\gamma}z^{\alpha}_{\check 1}z^{\beta}_{\check 2}z^{\gamma}_{\check 3})
(\tau ,\vec \sigma ),\nonumber \\
&&l^2(\tau ,\vec \sigma )=1,\quad\quad l_{\mu}(\tau ,\vec \sigma )z^{\mu}
_{\check r}(\tau ,\vec \sigma )=0,
\label{a7}
\end{eqnarray}

\noindent is the unit (future pointing) normal to $\Sigma_{\tau}$ at
$z^{\mu}(\tau ,\vec \sigma )$.

For the volume element in Minkowski spacetime we have

\begin{eqnarray}
d^4z&=&z^{\mu}_{\tau}(\tau ,\vec \sigma )d\tau d^3\Sigma_{\mu}=d\tau [z^{\mu}
_{\tau}(\tau ,\vec \sigma )l_{\mu}(\tau ,\vec \sigma )]\sqrt{\gamma
(\tau ,\vec \sigma )}d^3\sigma=\nonumber \\
&=&\sqrt{g(\tau ,\vec \sigma )} d\tau d^3\sigma.
\label{a8}
\end{eqnarray}

Let us remark that according to the geometrical approach of 
Ref.\cite{kuchar},one 
can use Eq.(\ref{a5}) in the form $z^{\mu}_{\tau}(\tau ,\vec \sigma )=N(\tau ,
\vec \sigma )l^{\mu}(\tau ,\vec \sigma )+N^{\check r}(\tau ,\vec \sigma )
z^{\mu}_{\check r}(\tau ,\vec \sigma )$, where $N=\sqrt{g/\gamma}=\sqrt{g
_{\tau\tau}-\gamma^{{\check r}{\check s}}g_{\tau{\check r}}g_{\tau{\check s}}}$ 
and $N^{\check r}=g_{\tau \check s}\gamma^{\check s\check r}$ are the 
standard lapse and shift functions, so that $g_{\tau \tau}=N^2+
g_{\check r\check s}N^{\check r}N^{\check s}, g_{\tau \check r}=
g_{\check r\check s}N^{\check s},
g^{\tau \tau}=N^{-2}, g^{\tau \check r}=-N^{\check r}/N^2, g^{\check r\check
s}=\gamma^{\check r\check s}+{{N^{\check r}N^{\check s}}\over {N^2}}$,
${{\partial}\over {\partial z^{\mu}_{\tau}}}=l_{\mu}\, {{\partial}\over
{\partial N}}+z_{{\check s}\mu}\gamma^{{\check s}{\check r}} {{\partial}\over
{\partial N^{\check r}}}$, $d^4z=N\sqrt{\gamma}d\tau d^3\sigma$.

The rest frame form of a timelike fourvector $p^{\mu}$ is $\stackrel
{\circ}{p}{}^{\mu}=\eta \sqrt{p^2} (1;\vec 0)= \eta^{\mu o}\eta \sqrt{p^2}$,
$\stackrel{\circ}{p}{}^2=p^2$, where $\eta =sign\, p^o$.
The standard Wigner boost transforming $\stackrel{\circ}{p}{}^{\mu}$ into
$p^{\mu}$ is

\begin{eqnarray}
L^{\mu}{}_{\nu}(p,\stackrel{\circ}{p})&=&\epsilon^{\mu}_{\nu}(u(p))=
\nonumber \\
&=&\eta^{\mu}_{\nu}+2{ {p^{\mu}{\stackrel{\circ}{p}}_{\nu}}\over {p^2}}-
{ {(p^{\mu}+{\stackrel{\circ}{p}}^{\mu})(p_{\nu}+{\stackrel{\circ}{p}}_{\nu})}
\over {p\cdot \stackrel{\circ}{p} +p^2} }=\nonumber \\
&=&\eta^{\mu}_{\nu}+2u^{\mu}(p)u_{\nu}(\stackrel{\circ}{p})-{ {(u^{\mu}(p)+
u^{\mu}(\stackrel{\circ}{p}))(u_{\nu}(p)+u_{\nu}(\stackrel{\circ}{p}))}
\over {1+u^o(p)} },\nonumber \\
&&{} \nonumber \\
\nu =0 &&\epsilon^{\mu}_o(u(p))=u^{\mu}(p)=p^{\mu}/\eta \sqrt{p^2}, \nonumber \\
\nu =r &&\epsilon^{\mu}_r(u(p))=(-u_r(p); \delta^i_r-{ {u^i(p)u_r(p)}\over
{1+u^o(p)} }).
\label{a9}
\end{eqnarray}

The inverse of $L^{\mu}{}_{\nu}(p,\stackrel{\circ}{p})$ is $L^{\mu}{}_{\nu}
(\stackrel{\circ}{p},p)$, the standard boost to the rest frame, defined by

\begin{equation}
L^{\mu}{}_{\nu}(\stackrel{\circ}{p},p)=L_{\nu}{}^{\mu}(p,\stackrel{\circ}{p})=
L^{\mu}{}_{\nu}(p,\stackrel{\circ}{p}){|}_{\vec p\rightarrow -\vec p}.
\label{a10}
\end{equation}

Therefore, we can define the following vierbeins [the $\epsilon^{\mu}_r(u(p))$'s
are also called polarization vectors; the indices r, s will be used for A=1,2,3
and $\bar o$ for A=0]

\begin{eqnarray}
&&\epsilon^{\mu}_A(u(p))=L^{\mu}{}_A(p,\stackrel{\circ}{p}),\nonumber \\
&&\epsilon^A_{\mu}(u(p))=L^A{}_{\mu}(\stackrel{\circ}{p},p)=\eta^{AB}\eta
_{\mu\nu}\epsilon^{\nu}_B(u(p)),\nonumber \\
&&{} \nonumber \\
&&\epsilon^{\bar o}_{\mu}(u(p))=\eta_{\mu\nu}\epsilon^{\nu}_o(u(p))=u_{\mu}(p),
\nonumber \\
&&\epsilon^r_{\mu}(u(p))=-\delta^{rs}\eta_{\mu\nu}\epsilon^{\nu}_r(u(p))=
(\delta^{rs}u_s(p);\delta^r_j-\delta^{rs}\delta_{jh}{{u^h(p)u_s(p)}\over
{1+u^o(p)} }),\nonumber \\
&&\epsilon^A_o(u(p))=u_A(p),
\label{a11}
\end{eqnarray}

\noindent which satisfy

\begin{eqnarray}
&&\epsilon^A_{\mu}(u(p))\epsilon^{\nu}_A(u(p))=\eta^{\mu}_{\nu},\nonumber \\
&&\epsilon^A_{\mu}(u(p))\epsilon^{\mu}_B(u(p))=\eta^A_B,\nonumber \\
&&\eta^{\mu\nu}=\epsilon^{\mu}_A(u(p))\eta^{AB}\epsilon^{\nu}_B(u(p))=u^{\mu}
(p)u^{\nu}(p)-\sum_{r=1}^3\epsilon^{\mu}_r(u(p))\epsilon^{\nu}_r(u(p)),
\nonumber \\
&&\eta_{AB}=\epsilon^{\mu}_A(u(p))\eta_{\mu\nu}\epsilon^{\nu}_B(u(p)),\nonumber 
\\
&&p_{\alpha}{{\partial}\over {\partial p_{\alpha}} }\epsilon^{\mu}_A(u(p))=
p_{\alpha}{{\partial}\over {\partial p_{\alpha}} }\epsilon^A_{\mu}(u(p))
=0.
\label{a12}
\end{eqnarray}

The Wigner rotation corresponding to the Lorentz transformation $\Lambda$ is

\begin{eqnarray}
R^{\mu}{}_{\nu}(\Lambda ,p)&=&{[L(\stackrel{\circ}{p},p)\Lambda^{-1}L(\Lambda
p,\stackrel{\circ}{p})]}^{\mu}{}_{\nu}=\left(
\begin{array}{cc}
1 & 0 \\
0 & R^i{}_j(\Lambda ,p)
\end{array}  
\right) ,\nonumber \\
{} && {}\nonumber \\
R^i{}_j(\Lambda ,p)&=&{(\Lambda^{-1})}^i{}_j-{ {(\Lambda^{-1})^i{}_op_{\beta}
(\Lambda^{-1})^{\beta}{}_j}\over {p_{\rho}(\Lambda^{-1})^{\rho}{}_o+\eta 
\sqrt{p^2}} }-\nonumber \\
&-&{{p^i}\over {p^o+\eta \sqrt{p^2}} }[(\Lambda^{-1})^o{}_j- { {((\Lambda^{-1})^o
{}_o-1)p_{\beta}(\Lambda^{-1})^{\beta}{}_j}\over {p_{\rho}(\Lambda^{-1})^{\rho}
{}_o+\eta \sqrt{p^2}} }].
\label{a13}
\end{eqnarray}

The polarization vectors transform under the 
Poincar\'e transformations $(a,\Lambda )$ in the following way

\begin{equation}
\epsilon^{\mu}_r(u(\Lambda p))=(R^{-1})_r{}^s\, \Lambda^{\mu}{}_{\nu}\, 
\epsilon^{\nu}_s(u(p)).
\label{a14}
\end{equation}

In Ref.\cite{lus1}, the system of N charged scalar 
particles was considered. As said in the Introduction, on the hypersurface
$\Sigma_{\tau}$ the particles are described by variables ${\vec \eta}_i(\tau )$
such that $x^{\mu}_i(\tau )=z^{\mu}(\tau ,{\vec \eta}_i(\tau ))$.
The electric charge of each particle is 
described in a pseudoclassical way\cite{casalb} by means of a pair of
complex conjugate Grassmann variables\cite{casala} $\theta_i(\tau ), \theta
^{*}_i(\tau )$ satisfying [$I_i=I^{*}_i=\theta^{*}_i\theta_i$ is the generator 
of the $U_{em}(1)$ group of particle i]

\begin{eqnarray}
&&\theta^2_i=\theta_i^{{*}2}=0,\quad\quad \theta_i\theta^{*}_i+\theta^{*}_i
\theta_i=0,\nonumber \\
&&\theta_i\theta_j=\theta_j\theta_i,\quad\quad \theta_i\theta^{*}_j=
\theta_j^{*}\theta_i,\quad\quad \theta^{*}_i\theta^{*}_j=\theta^{*}_j\theta
^{*}_i,\quad\quad i\not= j.
\label{a15}
\end{eqnarray}

\noindent This amounts to assume that the electric charges $Q_i=e_i\theta^{*}_i
\theta_i$ are quatized with levels 0 and $e_i$\cite{casala}. 

On the hypersurface $\Sigma_{\tau}$, we describe the electromagnetic potential
and field strength with Lorentz-scalar variables $A_{\check A}(\tau ,\vec \sigma
)$ and $F_{{\check A}{\check B}}(\tau ,\vec \sigma )$ respectively, defined by

\begin{eqnarray}
&&A_{\check A}(\tau ,\vec \sigma )=z^{\mu}_{\check A}(\tau ,\vec \sigma )
A_{\mu}(z(\tau ,\vec \sigma )),\nonumber \\
&&F_{{\check A}{\check B}}(\tau ,\vec \sigma )={\partial}_{\check A}A_{\check
B}(\tau ,\vec \sigma )-{\partial}_{\check B}A_{\check A}(\tau ,\vec \sigma )=
z^{\mu}_{\check A}(\tau ,\vec \sigma )z^{\nu}_{\check B}(\tau ,\vec \sigma )
F_{\mu\nu}(z(\tau ,\vec \sigma )).
\label{a16}
\end{eqnarray}

The system is described by the action\cite{lus1}

\begin{eqnarray}
S&=& \int d\tau d^3\sigma \, {\cal L}(\tau ,\vec
\sigma )=\int d\tau L(\tau ),\nonumber \\
L(\tau )&=&\int d^3\sigma {\cal L}(\tau ,\vec \sigma ),\nonumber \\
{\cal L}(\tau ,\vec \sigma )&=&{i\over 2}\sum_{i=1}^N\delta^3(\vec \sigma 
-{\vec \eta}_i(\tau ))[\theta^{*}_i(\tau ){\dot \theta}_i(\tau )-
{\dot \theta}^{*}_i(\tau )\theta_i(\tau )]-\nonumber \\
&&-\sum_{i=1}^N\delta^3(\vec \sigma -{\vec \eta}_i
(\tau ))[\eta_im_i\sqrt{ g_{\tau\tau}(\tau ,\vec \sigma )+2g_{\tau {\check r}}
(\tau ,\vec \sigma ){\dot \eta}^{\check r}_i(\tau )+g_{{\check r}{\check s}}
(\tau ,\vec \sigma ){\dot \eta}_i^{\check r}(\tau ){\dot \eta}_i^{\check s}
(\tau )  }+\nonumber \\
&&+e_i\theta^{*}_i(\tau )\theta_i(\tau )(A_{\tau }(\tau ,\vec \sigma )+
A_{\check r}(\tau ,\vec \sigma ){\dot \eta}^{\check r}_i(\tau ))]-\nonumber \\
&&-{1\over 4}\, \sqrt {g(\tau ,\vec \sigma )}
g^{{\check A}{\check C}}(\tau ,\vec \sigma )g^{{\check B}{\check D}}
(\tau ,\vec \sigma )F_{{\check A}{\check B}}(\tau ,\vec \sigma )
F_{{\check C}{\check D}}(\tau ,\vec \sigma ),
\label{a17}
\end{eqnarray}

\noindent where the configuration variables are $z^{\mu}(\tau ,\vec \sigma )$
$A_{\check A}(\tau ,\vec \sigma )$, ${\vec \eta}_i(\tau )$, $\theta_i(\tau )$
and $\theta^{*}_i(\tau )$, i=1,..,N. We have

$-{1\over 4}\sqrt{g} g^{\check A\check C}g^{\check B\check D}
\sum_aF_{a\check A
\check B}F_{a\check C\check D}=\hfill \break
=-{1\over 4}\sqrt{g}\sum_a [2(g^{\tau\tau}
g^{{\check r}{\check s}}-g^{\tau{\check r}}g^{\tau{\check s}}) 
F_{a\tau{\check r}}F_{a\tau{\check s}} +4g^{{\check r}{\check s}}
g^{\tau {\check u}}
F_{a\tau {\check r}}F_{a{\check s}{\check u}}+g^{{\check r}{\check u}}
g^{{\check s}{\check v}}F_{a{\check r}{\check s}}F_{a{\check u}{\check v}}]=
\hfill \break
=-\sqrt{\gamma}\sum_a[{1\over 2}\sqrt{ {{\gamma}\over
g} } F_{a\tau \check r}\gamma^{\check r\check s}F_{a\tau \check s}-
\sqrt{{{\gamma}
\over g}} g_{\tau \check v}\gamma^{\check v\check r}F_{a\check r\check s}
\gamma^{\check s\check u}F_{a\tau \check u}+{1\over 4}\sqrt{ {g\over {\gamma}}}
\gamma^{\check r\check s}F_{a\check r\check u}F_{a\check s\check v}(\gamma
^{\check
u\check v}+2{{\gamma}\over g}g_{\tau \check m}\gamma^{\check m\check u}
g_{\tau \check n}\gamma^{\check n\check v})]=-{{\sqrt{\gamma}}\over {2N}}
(F_{\tau \check r}-N^{\check u}F_{\check u\check r})\gamma^{\check r\check s}
(F_{\tau \check s}-N^{\check v}F_{\check v\check s})-{{N\sqrt{\gamma}}\over 4}
\gamma^{\check r\check s}\gamma^{\check u\check v}F_{\check r\check u}
F_{\check s\check v}.$

The action is invariant under separate
$\tau$- and $\vec \sigma$-reparametrizations, since $A_{\tau}(\tau ,\vec \sigma
)$ transforms as a $\tau$-derivative; moreover, it is invariant under the odd
phase transformations $\delta \theta_i\mapsto i\alpha \theta_i$, generated by
the $I_i$'s.

The canonical momenta are [$E_{\check r}=F_{{\check r}\tau}$ and $B_{\check r}
={1\over 2}\epsilon_{{\check r}{\check s}{\check t}}F_{{\check s}{\check t}}$ 
($\epsilon_{{\check r}{\check s}{\check t}}=\epsilon^{{\check r}{\check s}
{\check t}}$) are the electric and magnetic fields respectively; for
$g_{\check A\check B}\rightarrow \eta_{\check A\check B}$ one gets 
$\pi^{\check r}=-E_{\check r}=E^{\check r}$]

\begin{eqnarray}
\rho_{\mu}(\tau ,\vec \sigma )&=&-{ {\partial {\cal L}(\tau ,\vec \sigma )}
\over {\partial z^{\mu}_{\tau}(\tau ,\vec \sigma )} }=\sum_{i=1}^N\delta^3
(\vec \sigma -{\vec \eta}_i(\tau ))\eta_im_i\nonumber \\
&&{ {z_{\tau\mu}(\tau ,\vec \sigma )+z_{{\check r}\mu}(\tau ,\vec \sigma )
{\dot \eta}_i^{\check r}(\tau )}\over {\sqrt{g_{\tau\tau}(\tau ,\vec \sigma )+
2g_{\tau {\check r}}(\tau ,\vec \sigma ){\dot \eta}_i^{\check r}(\tau )+
g_{{\check r}{\check s}}(\tau ,\vec \sigma ){\dot \eta}_i^{\check r}(\tau ){\dot
\eta}_i^{\check s}(\tau ) }} }+\nonumber \\
&&+{ {\sqrt {g(\tau ,\vec \sigma )}}\over 4}[(g^{\tau \tau}z_{\tau \mu}+
g^{\tau {\check r}}z_{\check r\mu})(\tau ,\vec \sigma )g^{{\check A}{\check C}}
(\tau ,\vec \sigma )g^{{\check B}{\check D}}(\tau ,\vec \sigma )F_{{\check A}
{\check B}}(\tau ,\vec \sigma )F_{{\check C}{\check D}}(\tau ,\vec \sigma )
-\nonumber \\
&-&2[z_{\tau \mu}(\tau ,\vec \sigma )(g^{\check A\tau}g^{\tau \check C}
g^{{\check B}{\check D}}+g^{{\check A}{\check C}}g^{\check B\tau}g^{\tau 
\check D})(\tau ,\vec \sigma )+\nonumber \\
&+&z_{\check r\mu}(\tau ,\vec \sigma )
(g^{{\check A}{\check r}}g^{\tau {\check C}}+g^{{\check A}\tau}g^{{\check r}
{\check C}})(\tau ,\vec \sigma )g^{{\check B}{\check D}}
(\tau ,\vec \sigma )]F_{{\check A}{\check B}}(\tau ,\vec \sigma )
F_{{\check C}{\check D}}(\tau ,\vec \sigma )]=
\nonumber \\
&&=[(\rho_{\nu}l^{\nu})l_{\mu}+(\rho_{\nu}z^{\nu}_{\check r})\gamma^{{\check r}
{\check s}}z_{{\check s}\mu}](\tau ,\vec \sigma ),\nonumber \\
&&{}\nonumber \\
\pi^{\tau}(\tau ,\vec \sigma )&=&{ {\partial L}\over {\partial \partial_{\tau}
A_{\tau}(\tau ,\vec \sigma )} }=0,\nonumber \\
\pi^{\check r}(\tau ,\vec \sigma )&=&{ {\partial L}\over {\partial \partial
_{\tau}A_{\check r}(\tau ,\vec \sigma )} }=-{ {\gamma (\tau ,\vec \sigma )}
\over {\sqrt {g(\tau ,\vec \sigma )}} }\gamma^{{\check r}{\check s}}(\tau ,
\vec \sigma )(F_{\tau {\check s}}-g_{\tau {\check v}}\gamma^{{\check v}
{\check u}}F_{{\check u}{\check s}})(\tau ,\vec \sigma )=\nonumber \\
&&={ {\gamma (\tau ,\vec \sigma )}\over {\sqrt {g(\tau ,\vec \sigma )}} }
\gamma^{{\check r}{\check s}}(\tau ,\vec \sigma )(E_{\check s}(\tau ,\vec 
\sigma )+g_{\tau {\check v}}(\tau ,\vec \sigma )\gamma^{{\check v}{\check u}}
(\tau ,\vec \sigma )\epsilon_{{\check u}{\check s}{\check t}} B_{\check t}
(\tau ,\vec \sigma )),\nonumber \\
&&{}\nonumber \\
\kappa_{i{\check r}}(\tau )&=&-{ {\partial L(\tau )}\over {\partial {\dot
\eta}_i^{\check r}(\tau )} }=\nonumber \\
&=&\eta_im_i{ {g_{\tau {\check r}}(\tau ,{\vec \eta}_i(\tau ))+g_{{\check r}
{\check s}}(\tau ,{\vec \eta}_i(\tau )){\dot \eta}_i^{\check s}(\tau )}\over
{ \sqrt{g_{\tau\tau}(\tau ,{\vec \eta}_i(\tau ))+
2g_{\tau {\check r}}(\tau ,{\vec \eta}_i(\tau )){\dot \eta}_i^{\check r}(\tau )+
g_{{\check r}{\check s}}(\tau ,{\vec \eta}_i(\tau )){\dot \eta}_i^{\check r}
(\tau ){\dot \eta}_i^{\check s}(\tau ) }} }+\nonumber \\
&+&e_i\theta^{*}_i(\tau )\theta_i(\tau )A_{\check r}(\tau ,{\vec \eta}_i
(\tau )),\nonumber \\
&&{}\nonumber \\
\pi_{\theta \,i}(\tau )&=&{{\partial L(\tau )}\over {\partial {\dot \theta}_i
(\tau )}}=-{i\over 2}\theta^{*}_i(\tau )\nonumber \\
\pi_{\theta^{*} \, i}(\tau )&=&{{\partial L(\tau )}\over {\partial {\dot 
\theta}^{*}_i(\tau )}}=-{i\over 2}\theta_i(\tau ),
\label{a18}
\end{eqnarray}

\noindent and the following Poisson brackets are assumed

\begin{eqnarray}
&&\lbrace z^{\mu}(\tau ,\vec \sigma ),\rho_{\nu}(\tau ,{\vec \sigma}^{'}\rbrace
=-\eta^{\mu}_{\nu}\delta^3(\vec \sigma -{\vec \sigma}^{'}),\nonumber \\
&&\lbrace A_{\check A}(\tau ,\vec \sigma ),\pi^{\check B}(\tau ,\vec 
\sigma^{'} )\rbrace =\eta^{\check B}_{\check A}
\delta^3(\vec \sigma -\vec \sigma^{'}),\nonumber \\
&&\lbrace \eta^{\check r}_i(\tau ),\kappa_{j{\check s}}(\tau )\rbrace =-
\delta_{ij}\delta^{\check r}_{\check s},\nonumber \\
&&\lbrace \theta_i(\tau ),\pi_{\theta \, j}(\tau )\rbrace =-\delta_{ij},
\nonumber \\
&&\lbrace \theta^{*}_i(\tau ),\pi_{\theta^{*} \, j}(\tau )\rbrace =-\delta_{ij}.
\label{a19}
\end{eqnarray}

The Grassmann momenta give rise to the second class constraints $\pi_{\theta
\, i}+{i\over 2}\theta^{*}_i\approx 0$, $\pi_{\theta^{*}\, i}+{i\over 2}
\theta_i\approx 0$ [$\lbrace \pi_{\theta \, i}+{i\over 2}\theta^{*}_i,
\pi_{\theta^{*}\, j}+{i\over 2}\theta_j\rbrace =-i\delta_{ij}$]; $\pi
_{\theta \, i}$ and $\pi_{\theta^{*}\, i}$ are then eliminated with the
help of Dirac brackets

\begin{equation}
\lbrace A,B\rbrace {}^{*}=\lbrace A,B\rbrace -i[\lbrace A,\pi_{\theta \, i}+
{i\over 2}\theta^{*}_i\rbrace \lbrace \pi_{\theta^{*}\, i}+{i\over 2}
\theta_i,B\rbrace  +\lbrace A,\pi_{\theta^{*}\, i}+{i\over 2}
\theta_i \rbrace \lbrace \pi_{\theta \, i}+{i\over 2}\theta^{*}_i,B\rbrace ]
\label{a20}
\end{equation}

\noindent so that the remaining Grassmann variables have the fundamental
Dirac brackets [which we will still denote $\lbrace .,.\rbrace$ for the sake of
simplicity]

\begin{eqnarray}
&&\lbrace \theta_i(\tau ),\theta_j(\tau )\rbrace = \lbrace \theta_i^{*}(\tau ),
\theta_j^{*}(\tau )\rbrace =0,\nonumber \\
&&\lbrace \theta_i(\tau ),\theta_j^{*}(\tau )\rbrace =-i\delta_{ij}.
\label{a21}
\end{eqnarray}

We obtain four primary constraints

\begin{eqnarray}
&{\cal H}_{\mu}&(\tau ,\vec \sigma )= \rho_{\mu}(\tau ,\vec \sigma )-
l_{\mu}(\tau ,\vec \sigma )[T_{\tau\tau}(\tau ,\vec \sigma )+
\nonumber \\
&+&\sum_{i=1}^N\delta^3(\vec \sigma -{\vec \eta}_i(\tau ))\times
\nonumber \\
&\eta_i&\sqrt{ m^2_i-\gamma^{{\check r}{\check s}}(\tau ,\vec \sigma )
[\kappa_{i{\check r}}(\tau )-e_i\theta^{*}_i(\tau )\theta_i(\tau )A_{\check r}
(\tau ,\vec \sigma )][\kappa_{i{\check s}}(\tau ) -e_i\theta^{*}_i(\tau )
\theta_i(\tau )A_{\check s}(\tau ,\vec \sigma )]  }\, \, ]-\nonumber \\
&-&z_{{\check r}\mu}(\tau ,\vec \sigma )\gamma^{{\check r}{\check s}}
(\tau ,\vec \sigma )\lbrace
-T_{\tau \check s}(\tau ,\vec \sigma )+\sum_{i=1}^N\delta^3(\vec \sigma -
{\vec \eta}_i(\tau ))[\kappa_{i{\check s}}-e_i\theta_i^{*}(\tau )\theta_i
(\tau )A_{\check s}(\tau ,\vec \sigma )] \rbrace \approx 0,
\label{a22}
\end{eqnarray}

\noindent where

\begin{eqnarray}
T_{\tau \tau}(\tau ,\vec \sigma )
&=&-{1\over 2}({1\over {\sqrt {\gamma}} }\pi^{\check r}g_{{\check r}{\check s}}
\pi^{\check s}-{ {\sqrt {\gamma}}\over 2}\gamma^{{\check r}{\check s}}
\gamma^{{\check u}{\check v}}F_{{\check r}{\check u}}F_{{\check s}{\check v}})
(\tau ,\vec \sigma ),\nonumber \\
T_{\tau {\check s}}(\tau ,\vec \sigma )&=&-F_{{\check s}{\check t}}(\tau ,\vec 
\sigma )\pi^{\check t}(\tau ,\vec \sigma )=-\epsilon_{{\check s}{\check t}
{\check u}}\pi^{\check t}(\tau ,\vec \sigma )B_{\check u}(\tau ,\vec \sigma )
=\nonumber \\
&=&{[\vec \pi (\tau ,\vec \sigma )\times 
\vec B(\tau ,\vec \sigma )]}_{\check s},
\label{a23}
\end{eqnarray}

\noindent are the energy density and the Poynting vector respectively. We use
the notation $(\vec \pi \times \vec B)_{\check s}=(\vec E\times \vec B)
_{\check s}$ because it is consistent
with  $\epsilon_{{\check s}{\check t}{\check u}}\pi^{\check t}
B_{\check u}$ in the flat metric limit 
$g_{{\check A}{\check B}}\rightarrow \eta _{{\check A}
{\check B}}$; in this limit $T_{\tau\tau}\rightarrow {1\over 2}({\vec E}^2+
{\vec B}^2)$.

These constraints are first class \cite{lus1} and their existence implies that 
the description of the system is independent from the choice of the
foliation.

Since the canonical Hamiltonian is (we assume boundary conditions for the
electromagnetic potential such that all the surface terms can be neglected;
see Ref.\cite{lusa}) 

\begin{eqnarray}
H_c&=&-\sum_{i=1}^N\kappa_{i{\check r}}(\tau ){\dot \eta}_i^{\check r}(\tau )+
\int d^3\sigma [\pi^{\check A}(\tau ,\vec \sigma )\partial_{\tau}A_{\check A}
(\tau ,\vec \sigma )-\rho_{\mu}(\tau ,\vec \sigma )z^{\mu}_{\tau}(\tau ,\vec
\sigma )-{\cal L}(\tau ,\vec \sigma )]=\nonumber \\
&=&\int d^3\sigma [\partial_{\check r}(\pi^{\check r}(\tau ,\vec \sigma )
A_{\tau}(\tau ,\vec \sigma ))-A_{\tau}(\tau ,\vec \sigma )\Gamma
(\tau ,\vec \sigma )]=-\int d^3\sigma A_{\tau}(\tau ,\vec \sigma )
\Gamma (\tau ,\vec \sigma ),
\label{a24}
\end{eqnarray}

\noindent with

\begin{equation}
\Gamma(\tau ,\vec \sigma )=\partial_{\check r}\pi^{\check r}
(\tau ,\vec \sigma )-\sum_{i=1}
^Ne_i\theta^{*}_i(\tau )\theta_i(\tau )\delta^3(\vec \sigma -{\vec \eta}_i
(\tau )),
\label{a25}
\end{equation}

\noindent we have the Dirac Hamiltonian ($\lambda^{\mu}(\tau ,\vec \sigma )$ 
and $\lambda_{\tau}(\tau ,\vec \sigma )$  are Dirac's multipliers)

\begin{equation}
H_D=\int d^3\sigma \lambda^{\mu}(\tau ,\vec \sigma ){\cal H}_{\mu}(\tau ,\vec 
\sigma )+
\lambda_{\tau}(\tau ,\vec \sigma )\pi^{\tau}(\tau ,\vec \sigma )-
A_{\tau}(\tau ,\vec \sigma )\Gamma (\tau ,\vec \sigma )].
\label{a26}
\end{equation}

The Lorentz scalar constraint $\pi^{\tau}(\tau ,\vec \sigma )\approx 0$ is
generated by the gauge invariance of S; its time constancy will produce the
only secondary constraint (Gauss law) 

\begin{equation}
\Gamma (\tau ,\vec \sigma )\approx 0.
\label{a27}
\end{equation}

The ten conserved Poincar\'e generators are

\begin{eqnarray}
P^{\mu}&=&p^{\mu}_s=\int d^3\sigma \rho^{\mu}(\tau ,\vec \sigma ), \nonumber \\
J^{\mu\nu}&=&J^{\mu\nu}_s=\int d^3\sigma (z^{\mu}\rho^{\nu}-z^{\nu}\rho^{\mu})
(\tau ,\vec \sigma ).
\label{a28}
\end{eqnarray}

As shown in Ref.\cite{lus1}, one can restrict himself to foliations whose leaves
are spacelike hyperplanes with constant timelike normal $l^{\mu}$, by adding
the gauge-fixings $z^{\mu}(\tau ,\vec \sigma ) \approx x^{\mu}_s(\tau )+b^{\mu}
_{\check r}(\tau )\sigma^{\check r}$ [$b^{\mu}_{\check A}=\{ b^{\mu}_{\tau}=
l^{\mu}=\epsilon^{\mu}{}_{\alpha\beta\gamma}b_1^{\alpha}(\tau )b^{\beta}_2(\tau 
)b_3^{\gamma}(\tau ); b^{\mu}_{\check r}(\tau ) \}$ ($\partial_{\tau}l^{\mu}=0$)
is an orthonormal tetrad] and by going to Dirac brackets. In this way the
hypersurface degrees of freedom $z^{\mu}(\tau ,\vec \sigma )$, $\rho_{\mu}
(\tau ,\vec \sigma )$, are reduced to 20 ones: i) 8 are $x^{\mu}_s(\tau )$,
$p_s^{\mu}$; ii) 12 are the 6 independent pairs of canonical variables hidden in
$b^{\mu}_{\check A}$ and $S^{\mu\nu}_s=J^{\mu\nu}_s-(x^{\mu}_sp^{\nu}_s-x^{\nu}
_sp^{\mu}_s)$. The constraints ${\cal H}^{\mu}(\tau ,\vec \sigma )\approx 0$ are
reduced to only 10 constraints: ${\tilde {\cal H}}(\tau )=\int d^3\sigma {\cal
H}^{\mu}(\tau ,\vec \sigma ) \approx 0$, ${\tilde {\cal H}}(\tau )=b^{\mu}
_{\check r}(\tau ) \int d^3\sigma \, \sigma^{\check r} {\cal H}^{\nu}(\tau 
,\vec \sigma )-b^{\nu}_{\check r}(\tau ) \int d^3\sigma \, \sigma^{\check r} 
{\cal H}^{\mu}(\tau ,\vec \sigma ) \approx 0$.

Then, if one restricts himself to configurations with $p^2_s > 0$, one can 
make a further canonical reduction to the special foliation whose hyperplanes 
(the Wigner hyperplanes $\Sigma_{w\, \tau}$ are orthogonal to $p^{\mu}_s$ 
[namely $l^{\mu}=p^{\mu}_s/\sqrt{p_s^2}$]. This is achieved in two steps:
i) firstly, one boosts at rest the variables $b^{\mu}_{\check A}$, $S^{\mu\nu}
_s$, with the standard Wigner boost $L^{\mu}{}_{\nu}(p_s,{\buildrel \circ \over
p}_s)$ for timelike Poincar\'e orbits; ii) then, one adds the gauge-fixings
$b^{\mu}_{\check A}-L^{\mu}{}_a(p_s,{\buildrel \circ \over p}_s) \approx 0$
and goes to Dirac brackets. Only 4 pairs, ${\tilde x}^{\mu}_s(\tau )$, 
$p^{\mu}_s$, of canonical variables are associated with the Wigner hyperplane
$\Sigma_{w\, \tau}$ and the final constraints can be put in the form

\begin{eqnarray}
{\cal H}(\tau )&=&\eta_s\sqrt{p^2_s}-
[\sum_{i=1}^N\eta_i
\sqrt{m^2_i+[{\vec \kappa}_i(\tau )-e_i\theta^{*}_i(\tau )\theta_i(\tau )
\vec A(\tau ,{\vec \eta}_i(\tau ))]{}^2 }+\nonumber \\
&-&{1\over 2}\int d^3\sigma 
[{\vec \pi}^2(\tau ,\vec \sigma )+{\vec B}^2(\tau ,\vec \sigma ) ]
\approx 0,\nonumber \\
{\vec {\cal H}}_p(\tau )&=&
\sum_{i=1}^N[{\vec \kappa}_i(\tau )-e_i\theta
^{*}_i(\tau )\theta_i(\tau )\vec A(\tau ,{\vec \eta}_i(\tau ))]+
\nonumber \\
&+&\int d^3\sigma 
\vec \pi (\tau ,\vec \sigma )\times \vec B(\tau ,\vec \sigma )
\approx 0,\nonumber \\
\pi^{\tau}(\tau ,\vec \sigma )&\approx& 0,\nonumber \\
\Gamma (\tau ,\vec \sigma )&\approx& 0.
\label{a29}
\end{eqnarray}

As said in the Introduction, ${\tilde x}^{\mu}_s(\tau )$ is not a 4-vector
[see Ref.\cite{lus1} for its definition], 
$A_{\tau}(\tau ,\vec \sigma )$ and $\pi^{\tau}(\tau ,\vec \sigma )$ are
Lorentz scalars, while ${\vec A}(\tau ,\vec \sigma )$ and $\vec \pi (\tau 
,\vec \sigma )$ are Wigner spin-1 3-vectors.

As shown in Ref.\cite{lus1} and as it is shown in Section V for scalar 
electrodynamics, one can eliminate the electromagnetic gauge 
degrees of freedom and reexpress everything in terms of the Dirac observables:
i) ${\check {\vec A}}_{\perp}(\tau ,\vec \sigma )$, ${\check {\vec \pi}}
_{\perp}(\tau ,\vec \sigma )$, $\lbrace {\check A}^r_{\perp}(\tau ,\vec 
\sigma ),{\check \pi}^s_{\perp}(\tau ,{\vec 
\sigma}^{'})\rbrace =-P^{rs}_{\perp}(\vec \sigma )\delta^3(\vec \sigma -{\vec
\sigma}^{'})$ [$P^{rs}_{\perp}(\vec \sigma )=\delta^{rs}+{{\partial^r\partial^s}
\over {\triangle}}$, $\triangle =-{\vec \partial}^2$] for the electromagnetic
field; ii) ${\vec \eta}_i(\tau )$, ${\check {\vec \kappa}}_i(\tau )={\vec
\kappa}_i(\tau )+Q_i{{\vec \partial}\over {\triangle}}\, \vec \partial \cdot
\vec A(\tau ,\vec \sigma )$ for the particles [they become dressed with a 
Coulomb cloud]; iii) ${\check \theta}_i^{*}(\tau )$, ${\check \theta}_i(\tau )$,
such that $Q_i=e_i\theta_i^{*}\theta_i=e_i{\check \theta}_i^{*}{\check 
\theta}_i$.
This is the Wigner-covariant rest-frame Coulomb gauge. The reduced form of the
4 remaining constraints is

\begin{eqnarray}
{\cal H}(\tau )&=&\epsilon_s-\{ 
\sum_{i=1}^N\eta_{i}\sqrt{m_{i}^{2}+(\check{\vec{\kappa}}_{i}(\tau)
-Q_{i}\vec{A}_{\perp}(\tau,\vec{\eta}_{i}(\tau )))^{2}}
+\nonumber\\
&+&\sum_{i\neq j}\frac{Q_{i}Q_{j}}
{4\pi\mid\vec{\eta}_{i}(\tau )-\vec{\eta}_{j}(\tau )\mid}+
\int d^{3}\sigma\:
{1\over 2}[\check{\vec{\pi}}^{2}_{\perp}(\tau,\vec{\sigma})
+\check{\vec{B}}^{2}(\tau,\vec{\sigma})]\}\approx 0,\nonumber \\
{\vec {\cal H}}_p(\tau )&=&{\check {\vec \kappa}}_{+}(\tau )+\int d^3\sigma
[{\check {\vec \pi}}_{\perp}\times {\check {\vec B}}](\tau ,\vec \sigma
)\approx 0,
\label{a30}
\end{eqnarray}

Note that in this way one has extracted the Coulomb potential from field theory
and that the pseudoclassical property $Q^2_i=0$ regularizes the Coulomb 
self-energies.

\vfill\eject

\section{Reduced Hamilton Equations for the Electromagnetic Field and the 
Particles.}

Let us add some more results regarding N scalar electrically charged particles
plus the electromagnetic field in the rest-frame instant form of the dynamics
\cite{lus1}.

Let us first consider N scalar free particles, which are described by the
following 4 first class constraints on the Wigner hyperplane [on it the
independent Hamiltonian variables are: ($T_s$,$\epsilon_s$), ${\vec z}_s$,
${\vec k}_s$), (${\vec \eta}_i(\tau )$, ${\vec \kappa}_i(\tau )$), i=1,..,N;
the other basis for the particles is (${\vec \eta}_{+}(\tau )$, ${\vec \kappa}
_{+}(\tau )$), (${\vec \rho}_{\bar a}(\tau )$, ${\vec \pi}_{\bar a}(\tau )$),
$\bar a=1,..,N-1$]

\begin{eqnarray}
{\cal H}(\tau )&=& \epsilon_s -\sum_{i=1}^N\eta_{i}
\sqrt{m_{i}^{2}+\vec{\kappa}_{i}^{2}(\tau)}=\epsilon_s-H_{rel}
\approx 0,\nonumber\\
\vec{{\cal H}}_p(\tau)&=&\sum_{i}\vec{\kappa}_{i}(\tau)\approx 0,\nonumber \\
&&{}\nonumber \\
H_D&=&\lambda (\tau ) {\cal H}(\tau )-\vec \lambda (\tau )\cdot {\vec {\cal H}}_p
(\tau ).
\label{f1}
\end{eqnarray}

If we add the gauge-fixing

\begin{equation}
\chi=T_{s}-\tau\approx 0,
\label{f2}
\end{equation}

\noindent its conservation in $\tau$ will imply $\lambda (\tau )=-1$. Going to
Dirac brackets, we can eliminate the pair ($T_s$,$\epsilon_s$). The Hamiltonian
giving the evolution in the rest-frame time $\tau =T_s$ will be

\begin{eqnarray}
H_R&=&H_{rel}-\vec \lambda (\tau )\cdot {\vec {\cal H}}_p(\tau ),\nonumber \\
&&{}\nonumber \\
H_{rel}&=&\sum_{i=1}^N\eta_{i}\sqrt{m_{i}^{2}+\vec{\kappa}_{i}^{2}(\tau)}.
\label{f3}
\end{eqnarray}
 
\noindent The associated Hamilton-Dirac equations are

\begin{eqnarray}
\dot{\vec{\eta}}_{i}(\tau)\, &{\buildrel \circ \over =}&\, 
\frac{\vec{\kappa}_{i}(\tau)}{\eta_{i}
\sqrt{m_{i}^{2}+\vec{\kappa}^{2}_{i}(\tau)}}-\vec{\lambda}(\tau)\nonumber\\
\dot{\vec{\kappa}}_{i}(\tau)\, &{\buildrel \circ \over =}&\, 0,\nonumber \\
\sum_{i=1}^N {\vec \kappa}_i(\tau )\, &{\buildrel \circ \over =}&\, 0.
\label{f4}
\end{eqnarray}

\noindent The first line in invertible to give the momenta

\begin{equation}
\vec{\kappa}_{i}(\tau)=\eta_{i}m_{i}\frac{\dot{\vec{\eta}}_{i}(\tau)
+\vec{\lambda}(\tau)}{\sqrt{1-(\dot{\vec{\eta}}_{i}(\tau)
+\vec{\lambda}(\tau))^{2}}},
\label{f5}
\end{equation}

\noindent so that the associated Lagrangian is [$\vec \lambda (\tau )$ are
now Lagrange multipliers]

\begin{equation}
L_R=\vec{\kappa}_{i}(\tau )\cdot\dot{\vec{\eta}}_{i}(\tau )-H_R=
-\sum_{i=1}^N\eta_{i}m_{i}\, \sqrt{1-
(\dot{\vec{\eta}}_{i}(\tau)+\vec{\lambda}(\tau))^{2}}
\label{f6}
\end{equation}

\noindent The Euler-Lagrange equations are

\begin{eqnarray}
\frac{d}{d\tau}\frac{\partial L_R}{\partial\dot{\vec{\eta}}_{i}}&=&
\frac{d}{d\tau}(\eta_{i}m_{i}\frac{\dot{\vec{\eta}}_{i}(\tau)
+\vec{\lambda}(\tau)}{\sqrt{1
-(\dot{\vec{\eta}}_{i}(\tau)+\vec{\lambda}(\tau))^{2}}})
\, {\buildrel \circ \over =}\, 0\nonumber\\
\frac{\partial L_R}{\partial\vec{\lambda}}&=&\sum_{i}\eta_{i}m_{i}
\frac{\dot{\vec{\eta}}_{i}(\tau)
+\vec{\lambda}(\tau)}{\sqrt{1
-(\dot{\vec{\eta}}_{i}(\tau)+\vec{\lambda}(\tau))^{2}}}
\, {\buildrel \circ \over =}\, 0.
\label{f7}
\end{eqnarray}

Let us remark that there is no equation of motion for the variables (${\vec z}
_s$, ${\vec k}_s$) [the only left variables pertaining to the Wigner
hyperplane]: this means that they are Jacobi data independent from $\tau =T_s$.

In the free case one knows that by adding the gauge-fixings [its conservation
imply $\vec \lambda (\tau )=0$]

\begin{equation}
{\vec \eta}_{+}(\tau ) \approx 0,
\label{f8}
\end{equation}

\noindent one can eliminate, by going to Dirac brackets, the variables (${\vec 
\eta}_{+}$,${\vec \kappa}_{+}$). Now the N particles are descibed by N-1 pairs
of variables (${\vec \rho}_{\bar a}$,${\vec \pi}_{\bar a}$) relative to the
center of mass and there is no constraint left. The Hamiltonian for the
evolution in the rest-frame time $\tau =T_s$ is

\begin{equation}
H_{rel}=\sum_{i=1}^N\eta_{i}\sqrt{m^{2}_{i}+(N\sum_{\bar a=1}^{N-1}
\hat{\gamma}_{\bar ai}\check{\vec{\pi}}_{\bar a}(\tau))^{2}},
\label{f9}
\end{equation}

\noindent and the Hamilton equations are

\begin{eqnarray}
\dot{\vec{\rho}}_{a}(\tau )\, &{\buildrel \circ \over =}&\, 
\sum_{i=1}^N\sum_{\bar b=1}^{N-1}
\frac{N\hat{\gamma}_{ai}\hat{\gamma}_{bi}}{\eta_{i}
\sqrt{m^{2}_{i}+(N\sum_{\bar c=1}^{N-1}\hat{\gamma}_{\bar ci}{\vec{\pi}}
_{\bar c}(\tau))^{2}}}
{\vec{\pi}}_{\bar b},\nonumber\\
\dot{\vec{\pi}}_{\bar a}(\tau )\, &{\buildrel \circ \over =}&\, 0.
\label{f10}
\end{eqnarray}

\noindent However, this completely reduced description has the drawback that 
it is algebraically impossible to get explicitely the associated Lagrangian.
Only in the case N=2 with $m_1=m_2=m$ one gets

\begin{equation}
L_{rel}=-m\sqrt{4-\dot{\vec{\rho}}^{2}(\tau )},\,\,\, \Rightarrow |{\dot {\vec
\rho}}(\tau )| \leq 2.
\label{f11}
\end{equation} 

This result shows that there are kinematical restrictions on the relative
velocities, which, not being absolute velocities, can exceed the velocity of 
light without violation of Einstein causality.

Let us now study the case of N charged scalar particles plus the electromagnetic
field on the Wigner hyperplane after the addition of the gauge-fixing $T_s-\tau
\approx 0$. The Hamiltonian is\cite{lus1}

\begin{eqnarray}
H_R&=&H_{rel}-\vec{\lambda}(\tau)\cdot\vec{\cal H}_p(\tau),\nonumber \\
H_{rel}&=&\sum_{i=1}^N\eta_{i}\sqrt{m_{i}^{2}+(\check{\vec{\kappa}}_{i}(\tau)
-Q_{i}\vec{A}_{\perp}(\tau,\vec{\eta}_{i}(\tau )))^{2}}
+\nonumber\\
&+&\sum_{i\neq j}\frac{Q_{i}Q_{j}}
{4\pi\mid\vec{\eta}_{i}(\tau )-\vec{\eta}_{j}(\tau )\mid}+
\int d^{3}\sigma\:
{1\over 2}[\check{\vec{\pi}}^{2}_{\perp}(\tau,\vec{\sigma})
+\check{\vec{B}}^{2}(\tau,\vec{\sigma})],\nonumber \\
{\vec {\cal H}}_p(\tau )&=&{\check {\vec \kappa}}_{+}(\tau )+\int d^3\sigma
[{\check {\vec \pi}}_{\perp}\times {\check {\vec B}}](\tau ,\vec \sigma
)\approx 0,
\label{f12}
\end{eqnarray}

\noindent where ${\check {\vec \kappa}}_i$, ${\check {\vec A}}_{\perp}$,
${\check {\vec \pi}}_{\perp}$, are the electromagnetic Dirac observables.
The Grassmann-valued (Dirac observable) electric charges of the particles are
$Q_i=e_i{\check \theta}^{*}_i{\check \theta}_i$, $Q^2_i=0$.

The Hamilton-Dirac equations are

\begin{eqnarray}
\dot{\vec{\eta}}_{i}(\tau)\, &{\buildrel \circ \over =}&\,
\frac{
\check{\vec{\kappa}}_{i}(\tau)-Q_{i}\check{\vec{A}}_{\perp}(\tau,\vec{\eta}
_{i}(\tau ))}{\eta_i\sqrt{m_{i}^{2}+(
\check{\vec{\kappa}}_{i}(\tau)-Q_{i}\check{\vec{A}}_{\perp}(\tau,
\vec{\eta}_{i}(\tau )))^{2}}}
-\vec{\lambda}(\tau)\nonumber\\
\dot{\check{\vec{\kappa}}}_{i}(\tau)\, &{\buildrel \circ \over =}&\, 
-\sum_{k\neq i}\frac{Q_{i}Q_{k}({\vec \eta}_i(\tau )-{\vec \eta}_k(\tau )}
{4\pi\mid\vec{\eta}_{i}(\tau )-\vec{\eta}_{k}(\tau )\mid^{3}}+\nonumber\\
&+&Q_i(\dot{\eta}^{u}_{i}(\tau)+\lambda^{u}(\tau))
{{\partial}\over {{\vec \eta}_{i}}}{\check A}^{u}_{\perp}(\tau,\vec{\eta}_{i}
(\tau ))],\nonumber \\
{\check {\vec \kappa}}_{+}(\tau )&+&\int d^3\sigma
[{\check {\vec \pi}}_{\perp}\times {\check {\vec B}}](\tau ,\vec \sigma
)\approx 0.
\label{f13}
\end{eqnarray}

\noindent In the second equation we have already used the following inversion 
of the first equation

\begin{eqnarray}
\check{\vec{\kappa}}_{i}(\tau)&=&\eta_{i}
m_{i}\frac{\dot{\vec{\eta}}_{i}(\tau)+\vec{\lambda}(\tau)}
{\sqrt{1-
(\dot{\vec{\eta}}_{i}(\tau)+\vec{\lambda}(\tau))^{2}}}
+Q_{i}\check{\vec{A}}_{\perp}(\tau,\vec{\eta}_{i}(\tau )).
\label{f14}
\end{eqnarray}

\noindent The Hamilton-Dirac equations for the fields are

\begin{eqnarray}
\dot{A}_{\perp r}(\tau,\vec{\sigma})\, &{\buildrel \circ \over =}&\,
-\pi_{\perp r}(\tau,\vec{\sigma})
-[\vec{\lambda}(\tau)\cdot\vec{\partial}]A_{\perp r}(\tau,\vec{\sigma}),
\nonumber \\
\dot{\pi}^{r}_{\perp}(\tau,\vec{\sigma})\, &{\buildrel \circ \over =}&\,
\Delta A^{r}_{\perp}(\tau,\vec{\sigma})-
[\vec{\lambda}(\tau)\cdot\vec{\partial}]\pi^{r}_{\perp}(\tau,\vec{\sigma})
+\nonumber\\
&-&\sum_{i}Q_{i}P^{rs}_{\perp}(\vec{\sigma})\dot{\eta}_{i}^{s}(\tau)
\delta^3(\vec{\sigma}-\vec{\eta}_{i}(\tau )).
\label{f15}
\end{eqnarray}

\noindent The associated Lagrangian is

\begin{eqnarray}
L_R(\tau)&=&\dot{\vec{\eta}}_{i}(\tau)\cdot\check{\vec{\kappa}}_{i}(\tau)-
\int d^{3}\sigma\:{\dot {\check {\vec A}}}_{\perp}(\tau,\vec{\sigma})\cdot
\check{\vec{\pi}}_{\perp}(\tau,\vec{\sigma})-H_R(\tau)=\nonumber\\
&=&\sum_{i=1}^N[-\eta_{i}m_i\sqrt{1-
(\dot{\vec{\eta}}_{i}(\tau)+\vec{\lambda}(\tau))^{2}}+
Q_{i}[
\dot{\vec{\eta}}_{i}(\tau)+\vec{\lambda}(\tau)]\cdot\check{\vec{A}}_{\perp}
(\tau,\vec{\eta}_{i}(\tau ))]+\nonumber\\
&+&\frac{1}{2}\sum_{i\neq j}\frac{Q_{i}Q_{j}}
{4\pi\mid\vec{\eta}_{i}(\tau )-\vec{\eta}_{j}(\tau )\mid}+
\nonumber\\
&+&\int d^{3}\sigma\:[\frac{(\dot{\check{\vec{A}}}_{\perp}
+[\vec{\lambda}(\tau)\cdot\vec{\partial}]\check{\vec{A}}_{\perp})^{2}}{2}-
\frac{\check{\vec{B}}^{2}}{2}](\tau,\vec{\sigma})
\label{f16}
\end{eqnarray}

\noindent and its Euler-Lagrange equations are

\begin{eqnarray}
&&\frac{d}{d\tau}[\eta_{i}m_{i}\frac{\dot{\vec{\eta}}_{i}(\tau)
+\vec{\lambda}(\tau)}{\sqrt{1-(\dot{\vec{\eta}}_{i}(\tau)
+\vec{\lambda}(\tau))^{2}}}+Q_{i}\check{\vec{A}}_{\perp}(\tau,\vec{\eta}_{i}
(\tau ))]\, {\buildrel \circ \over =}\nonumber\\
&{\buildrel \circ \over =}\, &-\sum_{k\neq i}
\frac{Q_{i}Q_{k}({\vec \eta}_i(\tau )-{\vec \eta}_k(\tau )}
{4\pi\mid\vec{\eta}_{i}(\tau )-\vec{\eta}_{k}(\tau )\mid^{3}}
+Q_i(\dot{\eta}^{u}_{i}(\tau )+\lambda^u(\tau ))
{{\partial}\over {{\vec \eta}_{i}}}{\check A}^{u}_{\perp}(\tau,\vec{\eta}_{i}
(\tau )),
\label{f17}
\end{eqnarray}

\begin{eqnarray}
&-&\ddot{\check{A}}^{r}_{\perp}(\tau,\vec{\sigma})
-\frac{d}{d\tau}\{[\vec{\lambda}(\tau)\cdot\vec{\partial}]
{\check A}^{r}_{\perp}(\tau,\vec{\sigma})\} \, {\buildrel \circ \over =}
\nonumber \\
&{\buildrel \circ \over =}&\, \Delta {\check A}^{r}_{\perp}(\tau,\vec{\sigma})+
[\vec{\lambda}(\tau)\cdot\vec{\partial}]\{ \dot{\check{A}}^{r}_{\perp}(\tau,
\vec{\sigma})+[\vec{\lambda}(\tau)\cdot\vec{\partial}]{\check A}^{r}_{\perp}
(\tau,\vec{\sigma})\} +\nonumber \\
&-&\sum_{i=1}^NQ_iP_{\perp}^{rs}(\vec \sigma )[\dot{\eta}^{s}_{i}(\tau )+
\lambda^s(\tau )]\delta^3(\vec{\sigma}-\vec{\eta}_{i}(\tau )),
\label{f18}
\end{eqnarray}

\begin{eqnarray}
&&\sum_{i=1}^N[\eta_{i}m_i\frac{\dot{\vec{\eta}}_{i}(\tau)+\vec{\lambda}(\tau)}{
\sqrt{1-(\dot{\vec{\eta}}(\tau )+\vec{\lambda}(\tau))^{2}}}
+Q_{i}\check{\vec{A}}_{\perp}(\tau,\vec{\eta}_{i}(\tau ))]+\nonumber\\
&&+\int d^{3}\sigma\:
\sum_{r}[(\vec{\partial}{\check A}^{r}_{\perp})(\dot{\check{A}}^{r}_{\perp}+
[\vec{\lambda}(\tau)\cdot\vec{\partial}]{\check A}^{r}_{\perp})]
(\tau,\vec{\sigma})\, {\buildrel \circ \over =}\, 0.
\label{f19}
\end{eqnarray}

\noindent The Lagrangian expression for the invariant mass $H_{rel}$ is 

\begin{eqnarray}
E_{rel}&=&\sum_{i=1}^N\frac{\eta_{i}m_{i}}{\sqrt{1-(\dot{\vec{\eta}}_{i}(\tau)+
\vec{\lambda}(\tau))^{2}}}+\sum_{i>j}\frac{Q_{i}Q_{j}}
{4\pi\mid\vec{\eta}_{i}(\tau )-\vec{\eta}_{j}(\tau )\mid}+\nonumber \\
&+&\int d^{3}\sigma\: {1\over 2}[
                      \check{\vec{E}}_{\perp}^{2}(\tau,\vec{\sigma})+
                      \check{\vec{B}}^{2}(\tau,\vec{\sigma})]=const.
\label{f20}
\end{eqnarray}

\noindent Eq.(\ref{f17}) may be rewritten as

\begin{eqnarray}
&&\frac{d}{d\tau}(\eta_{i}m_{i}\frac{
\dot{\vec{\eta}}_{i}(\tau)+\vec{\lambda}(\tau)}{\sqrt{
1+(\dot{\vec{\eta}}_{i}(\tau)+\vec{\lambda}(\tau))^{2}}})\,
{\buildrel \circ \over =}\, \nonumber\\
&{\buildrel \circ \over =}&\, -\sum_{k\neq i}
\frac{Q_{i}Q_{k}({\vec \eta}_i(\tau )-{\vec \eta}_k(\tau ))}
{4\pi\mid\vec{\eta}_{i}(\tau )-\vec{\eta}_{k}(\tau )\mid^{3}}+\nonumber\\
&+&Q_{i}[\check{\vec{E}}_{\perp}(\tau,\vec{\eta}_i(\tau ))+
(\dot{\vec{\eta}}_{i}(\tau)+\vec{\lambda}(\tau))\times \check{\vec{B}}
(\tau,\vec{\eta}_i(\tau ))],
\label{f21}
\end{eqnarray}

\noindent where the notation ${\check E}^{r}_{\perp}=-\check{\dot{A}}^{r}
_{\perp}-[\vec{\lambda}(\tau)\cdot\vec{\partial}]{\check A}^{r}_{\perp}=
{\check \pi}^r_{\perp}$ has been introduced.

Let us remark that Eqs.(\ref{f21}) and (\ref{f18}) are the rest-frame analogues
of the usual equations for charged particles in an external electromagnetic 
field and of the electromagnetic field with external particle sources; however,
now, both particles and electromagnetic field are dynamical.  
 Eq.(\ref{f19}) defines the
rest frame by using the total (Wigner spin 1) 3-momentum of the isolated
system formed by the particles plus the electromagnetic field. Eq.(\ref{f20})
gives the constant invariant mass of the isolated system: the 
electromagnetic self-energy of the particles has been regularized by the
Grassmann-valued electric charges [$Q^2_i=0$] so that the invariant mass is
finite.

Let us now consider the gauge $\vec \lambda (\tau )=0$ [this result would
be consistent with the addition of the unknown gauge-fixings for the 3
first class constraints ${\vec {\cal H}}_p(\tau )\approx 0$].  In this case,
by eliminating the momenta, the equations of motion for the particles and
for the electromagnetic field become

\begin{eqnarray}
\frac{d}{d\tau}(\eta_{i}m_i\frac{\dot{\vec{\eta}}_{i}(\tau )}{\sqrt{1-
\dot{\vec{\eta}}_{i}^{2}(\tau )}})\, &{\buildrel \circ \over =}&\, -\sum_{k\neq
i}\frac{Q_{i}Q_{k}({\vec \eta}_i(\tau )-{\vec \eta}_k(\tau ))}
{4\pi\mid\vec{\eta}_{i}(\tau )-\vec{\eta}_{k}(\tau )\mid^{3}}+\nonumber\\
&+&Q_{i}[\check{\vec{E}}_{\perp}(\tau,\vec{\eta}_i(\tau ))+
\dot{\vec{\eta}}_{i}(\tau )\times \check{\vec{B}}(\tau,\vec{\eta}_i
(\tau ))],
\label{f22} 
\end{eqnarray}

\begin{eqnarray}
\Box {\check A}^r_{\perp}(\tau ,\vec \sigma )&=&
\ddot{\check A}^{r}_{\perp}(\tau,\vec{\sigma})+\Delta
{\check A}^{r}_{\perp}(\tau,\vec{\sigma})\, {\buildrel \circ \over =}\,J^{r}
_{\perp}(\tau,\vec{\sigma})=\nonumber\\
&=&\sum_{i=1}^NQ_{i}P^{rs}_{\perp}(\vec{\sigma})\dot{\eta}^{s}(\tau)
\delta^3(\vec{\sigma}-\vec{\eta}_{i}(\tau ))=\nonumber\\
&=&\sum_{i=1}^NQ_{i}\dot{\eta}^{s}(\tau)(\delta^{rs}+\frac{\partial^{r}
\partial^{s}}{\Delta})\delta^3(\vec{\sigma}-\vec{\eta}_{i}(\tau ))=\nonumber\\
&=&\sum_{i=1}^NQ_{i}\dot{\eta}^{s}(\tau)[\delta^3(\vec{\sigma}-\vec{\eta}_{i}
(\tau ))+\nonumber\\
&+&\int d^{3}\sigma'\:\frac{\pi^{rs}(\vec{\sigma}-\vec{\sigma}')}
{\mid\vec{\sigma}-\vec{\sigma}'\mid^{3}}\delta^3(\vec{\sigma}'-\vec{\eta}_{i}
(\tau )),
\label{f23}
\end{eqnarray}

\noindent with

\begin{equation}
\pi^{rs}(\vec{\sigma}-\vec{\sigma}')=\delta^{rs}-3
(\sigma^{r}-\sigma'^{r})(\sigma^{s}-\sigma'^{s})/(\vec{\sigma}-\vec{\sigma}')^
{2}.
\label{f24}
\end{equation}

Due to the projector $P^{rs}_{\perp}(\vec \sigma )$ required by the rest-frame
Coulomb gauge, the sources of the transverse (Wigner spin 1) vector potential
are no more local and one has a system of integrodifferential  equations 
(like with the equations generated by Fokker-Tetrode actions) with
the open problem of how to define an initial value problem.

The equations identifying the rest frame become

\begin{eqnarray}
&&\sum_{i=1}^N(\eta_{i}m_i\frac{\dot{\vec{\eta}}_{i}(\tau)}{
\sqrt{1-{\dot{\vec{\eta}}}^2(\tau )}}
+Q_{i}\check{\vec{A}}_{\perp}(\tau,\vec{\eta}_{i}(\tau )))+\nonumber\\
&&+\int d^{3}\sigma\:
\sum_{r}[(\vec{\partial}{\check A}^{r}_{\perp})(\dot{\check{A}}^{r}_{\perp}]
(\tau,\vec{\sigma})\, {\buildrel \circ \over =}\, 0.
\label{f25}
\end{eqnarray}

\vfill\eject

\section
{Electromagnetic Lienard-Wiechert Potentials.}

Eqs.(\ref{f23}) can be resolved by using the retarded Green function

\begin{eqnarray}
G_{ret}(\tau;\vec{\sigma})&=&(1/2\pi)\theta(\tau)\delta[\tau^{2}-
\vec{\sigma}^{2}],\nonumber \\
\Rightarrow && \Box G_{ret}(\tau,\vec{\sigma})=\delta(\tau)\delta^3
(\vec{\sigma}),
\label{f26}
\end{eqnarray}

and one obtains

\begin{eqnarray}
{\check A}^{r}_{\perp RET}(\tau,\vec{\sigma})\, &{\buildrel \circ \over =}&\,
{\check A}^r_{\perp IN}(\tau,\vec{\sigma})+
\sum_{i=1}^N\frac{Q_i}{2\pi}P_{\perp}^{rs}(\vec{\sigma})\int d\tau'
d^{3}\sigma'\:\nonumber\\
&&\theta(\tau-\tau')
\delta[(\tau-\tau')^{2}-(\vec{\sigma}-\vec{\sigma}')^{2}]\dot{\eta}^{s}_{i}
(\tau')\delta^3(\vec{\sigma}'-\vec{\eta}_{i}(\tau^{'}))=\nonumber\\
&=&{\check A}^r_{\perp IN}(\tau,\vec{\sigma})+\sum_{i=1}^N\frac{Q_i}{2\pi}
P_{\perp}^{rs}(\vec{\sigma})
\int d\tau'\:\nonumber\\
&&\theta(\tau-\tau')\delta[(\tau-\tau')^{2}-(\vec{\sigma}-\vec{\eta}_{i}(\tau
^{'}))^{2}]\dot{\eta}^{s}_{i}(\tau'),
\label{f27}
\end{eqnarray}

\noindent where $\Box A_{\perp IN}(\tau,\vec{\sigma})=0$ is a homogeneous
solution describing arbitrary incoming radiation.

Let $(\tau ,\vec \sigma )$ be the coordinates of a point $z^{\mu}(\tau ,\vec 
\sigma )$ of Minkowski spacetime lying on the Wigner hyperplane $\Sigma_W(\tau 
)$, on which the locations of the particles are $(\tau ,{\vec \eta}_i(\tau ))$
[i.e. $x^{\mu}_i(\tau )=z^{\mu}(\tau ,{\vec \eta}_i(\tau ))$]. The rest-frame
distance between $z^{\mu}(\tau ,\vec \sigma )$ and $x^{\mu}_i(\tau )$ is
${\vec r}_i(\tau ,\vec \sigma )=\vec \sigma -{\vec \eta}_i(\tau )$; let ${\hat
{\vec r}}_i(\tau ,\vec \sigma )=(\vec \sigma -{\vec \eta}_i(\tau ))/|\vec \sigma
-{\vec \eta}_i(\tau )|$ be the associated unit vector, ${\hat {\vec r}}_i^2
(\tau ,\vec \sigma )=1$.

Let $\tau_{i+}(\tau ,\vec \sigma )$ [the retarded times] denote the retarded 
solutions of the equations

\begin{equation} 
(\tau-\tau_{i+})^{2}=(\vec{\sigma}-\vec{\eta}_{i}(\tau_{i+}))^{2},\quad\quad
i=1,..,N.
\label{f28}
\end{equation}

\noindent The point $z^{\mu}(\tau ,\vec \sigma )$ lies on the lightcones 
emanating from the particle worldlines at their points $x^{\mu}_i(\tau_{i+}
(\tau ,\vec \sigma ))=z^{\mu}(\tau_{i+}(\tau ,\vec \sigma ),{\vec \eta}_i
(\tau_{i+}(\tau ,\vec \sigma )))$, lying on the Wigner hyperplanes $\Sigma_W
(\tau_{i+}(\tau ,\vec \sigma ))$ respectively. The point $z^{\mu}(\tau ,\vec 
\sigma )$ on $\Sigma_W(\tau )$ will define points $z^{\mu}(\tau_{i+}(\tau ,
\vec \sigma ),\vec \sigma )$ on the Wigner hyperplanes $\Sigma_W(\tau_{i+}
(\tau ,\vec \sigma ))$ by orthogonal projection [since $(z^{\mu}(\tau ,\vec 
\sigma )-x^{\mu}_i(\tau_{i+}(\tau ,\vec \sigma ))^2=0$, we have $R_{i+}
(\tau ,\vec \sigma )=\sqrt{(z(\tau ,\vec \sigma )-z(\tau_{i+}(\tau ,\vec 
\sigma ),\vec \sigma ))^2}=\sqrt{-(z(\tau_{i+}(\tau ,\vec \sigma ),\vec \sigma )
-x_i(\tau_{i+}(\tau ,\vec \sigma ))^2}$ and $z^{\mu}(\tau ,\vec \sigma )-x^{\mu}
_i(\tau_{i+}(\tau ,\vec \sigma ))=R_{i+}(\tau ,\vec \sigma )(t^{\mu}_{i+}
(\tau ,\vec \sigma )+s^{\mu}_{i+}(\tau ,\vec \sigma ))$ with $t^{\mu}_{i+}
(\tau ,\vec \sigma )$ and $s^{\mu}_{i+}(\tau ,\vec \sigma )$ being the timelike
and spacelike unit vectors associated with $z^{\mu}(\tau ,\vec \sigma )-
z^{\mu}(\tau_{i+}(\tau ,\vec \sigma ),\vec \sigma )$ and $z^{\mu}(\tau_{i+}
(\tau ,\vec \sigma ),\vec \sigma )-x^{\mu}_i(\tau_{i+}(\tau ,\vec \sigma ))$
respectively; $R_{i+}(\tau ,\vec \sigma )$ is the Minkowski retarded distance
between $z^{\mu}(\tau ,\vec \sigma )$ and $x^{\mu}_i(\tau_{i+}(\tau ,\vec 
\sigma ))$].

Let ${\vec r}_{i+}(\tau_{i+}(\tau ,\vec \sigma ),\vec \sigma )=\vec \sigma -
{\vec \eta}_i(\tau_{i+}
(\tau ,\vec \sigma ))$ denote the rest-frame retarded distance between the 
points $z^{\mu}(\tau_{i+}(\tau ,\vec \sigma ),\vec \sigma )$ and the points
$x^{\mu}_i(\tau_{i+}(\tau ,\vec \sigma ))$ of the worldlines belonging to 
$\Sigma_W(\tau_{i+}(\tau ,\vec \sigma ))$ [with ${\hat {\vec r}}_{i+}
(\tau_{i+}(\tau ,\vec \sigma ),\vec \sigma )$ being the unit vector, ${\hat 
{\vec r}}^2_{i+}=1$]. Let us denote the lenght of the vectors ${\vec r}_{i+}
(\tau_{i+}(\tau ,\vec \sigma ),\vec \sigma )$ with

\begin{eqnarray}
r_{i+}(\tau_{i+}(\tau ,\vec \sigma ),\vec \sigma )&=&|{\vec r}_{i+}(\tau
_{i+}(\tau ,\vec \sigma ),\vec \sigma )|=|\vec \sigma
-{\vec \eta}_i(\tau_{i+}(\tau ,\vec \sigma ))|=\nonumber \\
&=&\tau -\tau_{i+}(\tau ,\vec \sigma ) > 0.
\label{f29}
\end{eqnarray}

Then, we have

\begin{eqnarray}
\theta (\tau -\tau^{'})&& \delta [(\tau -\tau^{'})^2-(\vec \sigma -{\vec \eta}_i
(\tau^{'}))^2]={{\delta (\tau^{'}-\tau_{i+}(\tau ,\vec \sigma ))}\over
{2 \rho_{i+}(\tau_{i+}(\tau ,\vec \sigma ),\vec \sigma )}},\nonumber \\
&&{}\nonumber \\
\rho_{i+}(\tau_{i+}(\tau ,\vec \sigma ),\vec \sigma )&=&\tau -\tau_{i+}(\tau ,
\vec \sigma )-{\dot {\vec \eta}}_i(\tau_{i+}(\tau ,\vec \sigma ))\cdot [\vec 
\sigma -{\vec \eta}_i(\tau_{i+}(\tau ,\vec \sigma ))]=\nonumber \\
&=&r_{i+}(\tau_{i+}(\tau ,\vec \sigma ),\vec \sigma ) [1-{\dot {\vec \eta}}_i
(\tau_{i+}(\tau ,\vec \sigma ))\cdot {\hat {\vec r}}_{i+}(\tau_{i+}(\tau ,\vec 
\sigma ),\vec \sigma )].
\label{f30}
\end{eqnarray}

Eq.(\ref{f27}) can be rewritten as

\begin{eqnarray}
{\check A}^{r}_{\perp RET}(\tau,\vec{\sigma})&=&{\check A}^{r}_{\perp IN}(\tau,
\vec{\sigma})+\sum_{i=1}^N\frac{Q_{i}}{4\pi}P^{rs}_{\perp}(\vec{\sigma})
\frac{{\dot \eta}^{s}_{i}(\tau_{i+}(\tau ,\vec \sigma ))}{\varrho_{i+}
(\tau_{i+}(\tau ,\vec \sigma ),\vec \sigma )}=\nonumber\\
&=&{\check A}^{r}_{\perp IN}(\tau,\vec{\sigma})+\sum_{i=1}^N
\frac{Q_{i}}{4\pi}[\, \frac{{\dot \eta}^{s}_{i}(\tau_{i+}(\tau ,\vec \sigma ))}
{\varrho_{i+}(\tau_{i+}(\tau ,\vec \sigma ),\vec \sigma )}+\nonumber\\
&+&\int d^{3}\sigma'\frac{\pi^{rs}(\vec{\sigma}-\vec{\sigma}')}
{\mid\vec{\sigma}-\vec{\sigma}'\mid^{3}}\frac
{{\dot \eta}^{s}_{i}(\tau_{i+}(\tau,\vec{\sigma'}))}{\varrho_{i+}(\tau_{i+}
(\tau ,\vec \sigma^{'}),\vec{\sigma}')}\, ]=\nonumber\\
&=&{\check A}^{r}_{\perp IN}(\tau,\vec{\sigma})+\sum_{i=1}^N\check{A}^{r}
_{\perp (i+)}(\tau_{i+}(\tau ,\vec \sigma ),\vec{\sigma})=\nonumber \\
&=&{\check A}^{r}_{\perp IN}(\tau,\vec{\sigma})+\sum_{i=1}^NQ_i\tilde{A}^{r}
_{\perp (i+)}(\tau_{i+}(\tau ,\vec \sigma ),\vec{\sigma}),
\label{f31}
\end{eqnarray}

\noindent where ${\check {\vec A}}_{\perp (i+)}(\tau_{i+}(\tau ,\vec \sigma )
,\vec \sigma )$ is the 
rest-frame form of the Lienard-Wiechert retarded potential produced by particle 
i [its Minkowski analogue, i.e. the relativistic generalization of the Coulomb
potential, is $A^{\mu}_{(i+)}(z)={{Q_i}\over {4\pi}} {{{\dot x}^{\mu}_i(\tau
_{i+})}\over {{\dot x}_i(\tau_{i+})\cdot [z-x_i(\tau_{i+})]}}={{Q_i}\over 
{4\pi}} {{{\dot x}^{\mu}_i(\tau_{i+})}\over {R_{i+}[1-{\hat {\vec R}}_{i+}\cdot
{\dot {\vec x}}_i(\tau_{i+})]}}$]. Since we are in the rest-frame Coulomb gauge
with only transverse Wigner-covariant vector potentials, ${\check {\vec A}}
_{\perp (i+)}(\tau_{i+}(\tau ,\vec \sigma ),\vec \sigma )$ has a first standard 
term generated at the retarded time $\tau_{i+}(\tau ,\vec \sigma )$ at $x^{\mu}
_i(\tau_{i+}(\tau ,\vec \sigma ))$, which is, however, accompanied by a 
nonlocal term receiving contributions from all the retarded times $-\infty < 
\tau_{i+}(\tau ,{\vec \sigma}^{'}) \leq \tau$., which is due to the elimination
of the electromagnetic gauge degrees of freedom [this is the origin of the
transverse projector]. If we put the ${\check {\vec A}}_{\perp (i+)}(\tau_{i+}
(\tau ,\vec \sigma ),\vec \sigma )$'s in the particle equations (\ref{f22}), 
with ${\check {\vec A}}_{\perp IN}(\tau ,\vec \sigma )=0$, then the equations 
of motion become integro-differential equations like the ones generated by 
a Fokker action.

To evaluate the electric [${\check {\vec E}}_{\perp}=-{\dot {\check {\vec A}}}
_{\perp}$] and magnetic [${\check {\vec B}}=-\vec \partial \times
{\check {\vec A}}_{\perp}$] fields produced by ${\check {\vec A}}_{\perp (i+)}
(\tau ,\vec \sigma )$, we need the rule of derivation of `retarded' functions
$g(\tau ,\vec \sigma; \tau_{i+}(\tau ,\vec \sigma ))$. From Eq.(\ref{f28})
we get $(\tau -\tau_{i+})(d\tau -d\tau_{i+})=
r_{i+}(d\tau -d\tau_{i+})=[\vec \sigma -{\vec \eta}_i(\tau_{i+})]\cdot [d\vec 
\sigma -{\dot {\vec \eta}}_i(\tau_{i+})d\tau_{i+}]={\vec r}_{i+}\cdot
[d\vec \sigma -{\dot {\vec \eta}}_i(\tau_{i+})d\tau_{i+}]$. Therefore, by
introducing the notation

\begin{eqnarray}
\vec{v}_{i+}(\tau_{i+}(\tau ,\vec \sigma ),\vec \sigma )&=&\frac{\vec{r}_{i+}
(\tau_{i+}(\tau ,\vec \sigma ),\vec \sigma )}{\varrho_{i+}(\tau_{i+}(\tau ,\vec 
\sigma ),\vec \sigma )}=\frac{\hat{\vec{r}}_{i+}(\tau_{i+}(\tau ,\vec \sigma ),
\vec \sigma )}{1-\dot{\vec{\eta}}_{i}(\tau_{i+}(\tau ,\vec \sigma ))\cdot
\hat{\vec{r}}_{i+}(\tau_{i+}(\tau ,\vec \sigma ),\vec \sigma )},
\quad {\hat {\vec r}}_{i+}={{{\vec v}_{i+}}\over {|{\vec v}_{i+}|}},\nonumber \\
\:\mid\vec{v}_{i+}(\tau_{i+}(\tau ,\vec \sigma ),\vec \sigma )\mid &=& \frac{1}
{1-\dot{\vec{\eta}}{i}(\tau_{i+}(\tau ,\vec \sigma ))\cdot\hat{\vec{v}}_{i+}
(\tau_{i+}(\tau ,\vec \sigma ),\vec \sigma )}=\frac{\tau-\tau_{i+}(\tau ,\vec 
\sigma )}{\varrho_{i+}(\tau_{i+}(\tau ,\vec \sigma ),\vec \sigma )},
\label{f32}
\end{eqnarray}

\noindent we get

\begin{eqnarray}
{{\partial \tau_{i+}(\tau ,\vec \sigma )}\over {\partial \tau}}&=&|{\vec v}
_{i+}(\tau_{i+}(\tau ,\vec \sigma ),\vec \sigma )|,\nonumber \\
{{\partial \tau_{i+}(\tau ,\vec \sigma )}\over {\partial \sigma^s}}&=&{\hat r}
_{i+\, s}(\tau_{i+}(\tau ,\vec \sigma ),\vec \sigma )\, |{\vec v}_{i+}(\tau
_{i+}(\tau ,\vec \sigma ),\vec \sigma )|=v_{i+\, s}(\tau_{i+}(\tau ,\vec 
\sigma ),\vec \sigma ),\nonumber \\
{{\partial g(\tau ,\vec \sigma ;\tau_{i+}(\tau ,\vec \sigma )))}\over {\partial
\tau}}&=&[({{\partial}\over {\partial \tau}}{|}_{\tau^{'}}+|{\vec v}_{i+}
(\tau_{i+}(\tau ,\vec \sigma ),\vec \sigma )|{{\partial}\over {\partial \tau
^{'}}})g(\tau ,\vec \sigma ;\tau^{'})]{|}_{\tau^{'}=\tau_{i+}(\tau ,\vec 
\sigma )},\nonumber \\
{{\partial g(\tau ,\vec \sigma ;\tau_{i+}(\tau ,\vec \sigma )))}\over {\partial
\sigma^s}}&=&[({{\partial}\over {\partial \sigma^s}}{|}_{\tau^{'}}+v_{i+\, s}
(\tau_{i+}(\tau ,\vec \sigma ),\vec \sigma ){{\partial}
\over {\partial \tau^{'}}})g(\tau ,\vec \sigma ;\tau^{'})]{|}_{\tau^{'}=
\tau_{i+}(\tau ,\vec \sigma )}.
\label{f33}
\end{eqnarray}

\noindent so that [using $\partial {\vec r}_{i+}(\tau_{i+},\vec \sigma )/
\partial \tau_{i+}=-{\dot {\vec \eta}}_i(\tau_{i+})$, $\partial r_{i+}(\tau_{i+}
,\vec \sigma )/\partial \tau_{i+}=-{\dot {\vec \eta}}_i(\tau_{i+})\cdot {\hat 
{\vec r}}_{i+}(\tau_{i+},\vec \sigma )$, $\partial \rho_{i+}(\tau_{i+},\vec 
\sigma )/\partial \tau_{i+}={\dot {\vec \eta}}_i^2(\tau_{i+})-(\, {\dot {\vec 
\eta}}_i(\tau_{i+})+r_{i+}(\tau_{i+},\vec \sigma ){\ddot {\vec \eta}}_i(\tau
_{i+})\, )\cdot{\hat {\vec r}}_{i+}(\tau_{i+},\vec \sigma )$, ${{\partial}\over 
{\partial \sigma^s}}{|}_{\tau_{i+}}\, r^r_{i+}(\tau_{i+},\vec \sigma )=\delta
^r_s$, ${{\partial}\over {\partial \sigma^s}}{|}_{\tau_{i+}}\, r_{i+}(\tau
_{i+},\vec \sigma )={\hat r}^s_{i+}(\tau_{i+},\vec \sigma )$, ${{\partial}\over
{\partial \sigma^s}}{|}_{\tau_{i+}}\, \rho_{i+}(\tau_{i+},\vec \sigma )=-({\dot
\eta}_i^s(\tau_{i+})+{\hat r}^s_{i+}(\tau_{i+},\vec \sigma )\, )$] we get

\begin{eqnarray}
{\check E}^{r}_{\perp RET}(\tau ,\vec \sigma )&=&-{{\partial}\over {\partial 
\tau}}{\check A}^r_{\perp RET}(\tau_{i+}(\tau ,\vec \sigma ),\vec \sigma )
\,{\buildrel \circ \over =}\,
{\check E}^{r}_{\perp IN}(\tau,\vec{\sigma})-\nonumber\\
&-&P^{rs}_{\perp}(\vec{\sigma})\sum_{i=1}^N[\frac{Q_{i}}{4\pi}
|{\vec v}_{i+}(\tau_{i+}(\tau ,\vec \sigma ),\vec \sigma )|\, [{{{\ddot \eta}
^s_i(\tau_{i+}(\tau ,\vec \sigma )}\over {\rho_{i+}(\tau_{i+}(\tau ,\vec 
\sigma ),\vec \sigma )}}-\nonumber \\
&-&{{{\dot \eta}^s_i(\tau_{i+}(\tau ,\vec \sigma ))}\over {\rho^2_{i+}(\tau ,
\vec \sigma ),\vec \sigma )}} (\, {\dot {\vec \eta}}^2_i(\tau_{i+}(\tau ,\vec 
\sigma ))-\nonumber \\
&-&({\dot {\vec \eta}}_i(\tau_{i+}(\tau ,\vec \sigma ))+r_{i+}(\tau
_{i+}(\tau ,\vec \sigma ),\vec \sigma ) {\ddot {\vec \eta}}_i(\tau_{i+}(\tau ,
\vec \sigma ))\, )\cdot {\hat {\vec r}}_{i+}(\tau_{i+}(\tau ,\vec \sigma ),
\vec \sigma ))\, ]=\nonumber \\
&=&{\check E}^r_{\perp IN}(\tau ,\vec \sigma )+\sum_{i=1}^N{\check E}^r_{\perp 
(i+)}(\tau_{i+}(\tau ,\vec \sigma ),\vec \sigma )=\nonumber \\
&=&{\check E}^{r}_{\perp IN}(\tau,\vec{\sigma})+\sum_{i=1}^NQ_{i}\tilde{E}
^{r}_{\perp (i+)}(\tau_{i+}(\tau ,\vec \sigma ),\vec{\sigma}),
\label{f34}
\end{eqnarray}

\begin{eqnarray}
{\check B}^{r}_{RET}(\tau,\vec{\sigma})&=&-\epsilon^{rsu}(\partial^{s}
{\check A}^{u}_{\perp RET}(\tau,\vec{\sigma}))\, {\buildrel \circ \over =}\,
={\check B}^{r}_{IN}(\tau,\vec{\sigma})+\nonumber \\
&+&\sum_{i=1}^N{{Q_i}\over {4\pi}} \epsilon^{rsu}P^{uv}_{\perp}(\vec \sigma )
({{\partial}\over {\partial \sigma^s}}{|}_{\tau_{i+}}+v_{i+\, s}(\tau_{i+}
(\tau ,\vec \sigma ),\vec \sigma ){{\partial}\over {\partial \tau_{i+}}})
{{{\dot \eta}_i^v(\tau_{i+}(\tau ,\vec \sigma ))}\over {\rho_{i+}(\tau_{i+}
(\tau ,\vec \sigma ),\vec \sigma )}}=\nonumber \\
&=&{\check B}^r_{IN}(\tau ,\vec \sigma )+\sum_{i=1}^N [{\hat {\vec r}}_{i+}
(\tau_{i+}(\tau ,\vec \sigma ),\vec \sigma ) \times {\check {\vec E}}_{\perp 
(i+)}(\tau_{i+}(\tau ,\vec \sigma ),\vec \sigma )\, ]^r+\nonumber \\
&+&\sum_{i=1}^N{{Q_i}\over {4\pi}} \epsilon^{rsu}P^{uv}_{\perp}(\vec \sigma )
{{{\hat r}^s_{i+}(\tau_{i+}(\tau ,\vec \sigma ),\vec \sigma ) {\dot \eta}_i^v
(\tau_{i+}(\tau ,\vec \sigma ))}\over {\rho^2_{i+}(\tau_{i+}(\tau ,\vec \sigma )
,\vec \sigma )}}=\nonumber \\
&=&{\check B}^r_{IN}(\tau ,\vec \sigma )+\sum_{i=1}^N{\check B}^r_{(i+)}(\tau
_{i+}(\tau ,\vec \sigma ),\vec \sigma )=\nonumber \\
&=&{\check B}^r_{IN}(\tau ,\vec \sigma )+\sum_{i=1}^NQ_i{\tilde B}^r_{(i+)}
(\tau_{i+}(\tau ,\vec \sigma ),\vec \sigma ).
\label{f35}
\end{eqnarray}

\noindent The particle equations of motion (\ref{f21}), the definition of the
rest frame (\ref{f19}) and the conserved relative energy (\ref{f20}) have now
the following form 

\begin{eqnarray}
\frac{d}{d\tau}(\eta_{i}m_{i}\frac{\dot{\vec{\eta}}_{i}(\tau)}{\sqrt{
1-\dot{\vec{\eta}}_{i}^{2}(\tau)}})\, &{\buildrel \circ \over =}&\, -\sum
_{k\neq i}\frac{Q_{i}Q_{k}[{\vec \eta}_i(\tau )-{\vec \eta}_k(\tau )]}
{4\pi\mid\vec{\eta}_{i}(\tau )-\vec{\eta}_{k}(\tau )\mid^{3}}
+\nonumber\\
&+&Q_{i}[\check{\vec{E}}_{\perp IN}(\tau,\vec{\eta}_{i}(\tau ))+
\dot{\vec{\eta}}_{i}(\tau)\times\check{\vec{B}}_{IN}(\tau,\vec{\eta}_{i}(\tau )
)]+\nonumber\\
&+&\sum_{k\neq i}Q_{i}Q_{k}[\vec{\tilde{E}}_{\perp (k+)}(\tau_{i+}(\tau ,
{\vec \eta}_i(\tau )),\vec{\eta}_{i}(\tau ))+\nonumber \\
&+&\dot{\vec{\eta}}_{i}(\tau)\times
\vec{\tilde{B}}_{(k+)}(\tau_{i+}(\tau ,{\vec \eta}_i(\tau )),\vec{\eta}_{i}
(\tau ))],\nonumber \\
&&{}\nonumber \\
&&\sum_{i=1}^N[\eta_{i}m_{i}\frac{\eta_{i}m_{i}}{\sqrt{
1-\dot{\vec{\eta}}^{2}_{i}(\tau)}}+Q_{i}{\check {\vec A}}_{\perp IN}(\tau,
\vec{\eta}_{i}(\tau ))]+\nonumber \\
&+&\sum_{i\neq j}Q_{i}Q_{j}\vec{\tilde{A}}_{\perp (j+)}(\tau_{j+}(\tau ,{\vec 
\eta}_i(\tau ),\vec{\eta}_{i}(\tau ))+\nonumber\\
&+&\int d^{3}\sigma\:
[\check{\vec{E}}_{\perp IN}\times \check{\vec{B}}_{IN}
+\sum_{i=1}^NQ_{i}(\check{\vec{E}}_{\perp IN}\times\vec{\tilde{B}}_{\perp (i+)}
(\tau_{i+},\vec \sigma )+\nonumber \\
&+&\vec{\tilde{E}}_{\perp (i+)}(\tau_{i+},\vec \sigma )
\times\check{\vec{B}}_{IN})+\nonumber\\
&+&\sum_{i\neq j}Q_{i}Q_{j}\vec{\tilde{E}}_{\perp (i+)}(\tau_{i+},\vec \sigma )
\times\vec{\tilde{B}}_{(j+)}(\tau_{j+},\vec \sigma )](\tau,\vec{\sigma})\,
{\buildrel \circ \over =}\, 0,\nonumber \\
&&{}\nonumber \\    
E_{rel}&=&\sum_{i}\frac{\eta_{i}m_{i}}{\sqrt{
1-\dot{\vec{\eta}}_{i}^{2}(\tau)}}+\sum_{i\neq j}\frac{Q_{i}Q_{j}}
{4\pi\mid\vec{\eta}_{i}(\tau )-\vec{\eta}_{j}(\tau )\mid}+\nonumber\\
&+&\int d^{3}\sigma\:[
\frac{\check{\vec{E}}^{2}_{\perp IN}+\check{\vec{B}}^{2}_{IN}}{2}
+\nonumber\\
&+&\sum_{i=1}^NQ_{i}(\check{\vec{E}}_{\perp IN}\cdot\vec{\tilde{E}}_{\perp
(i+)}(\tau_{i+},\vec \sigma )+\check{\vec{B}}_{IN}\cdot\vec{\tilde{B}}_{(i+)}
(\tau_{i+},\vec \sigma ))+\nonumber\\
&+&\sum_{i>j}Q_{i}Q_{j}(\vec{\tilde{E}}_{\perp (i+)}(\tau_{i+},\vec \sigma )
\cdot\vec{\tilde{E}}_{\perp (j+)}(\tau_{j+},\vec \sigma )+\nonumber \\
&+&\vec{\tilde{B}}_{(i+)}(\tau_{i+},\vec \sigma )\cdot\vec{\tilde{E}}_{\perp 
(j+)}(\tau_{j+},\vec \sigma ))\, ](\tau,\vec{\sigma})=const.
\label{f36}
\end{eqnarray}
  
The property $Q^2_i=0$ has been used in these equations and it will be used also
in what follows. Besides the divergent Coulomb self-interaction it eliminates
other divergent terms. By using the equations of motion, it is verified that
$E_{rel}$ is a constant of the motion

\begin{eqnarray}
\frac{d}{d\tau}E_{rel}&=&\frac{d}{d\tau}[\sum_{i=1}^N\frac{\eta_{i}m_{i}}{\sqrt{
1-\dot{\vec{\eta}}_{i}^{2}(\tau )}}-\sum_{i>j}\frac{Q_{i}Q_{j}}
{4\pi\mid\vec{\eta}_{i}(\tau )-\vec{\eta}_{j}(\tau )\mid}]+\nonumber\\
&-&\sum_{i=1}^NQ_{i}\dot{\vec{\eta}}_{i}(\tau)\cdot\check{\vec{E}}_{\perp IN}
(\tau,\vec{\eta}_{i}(\tau ))-\sum_{i\neq j}Q_{i}Q_{j}
\dot{\vec{\eta}}_{i}(\tau)\cdot\vec{\tilde{E}}_{\perp (j+)}(\tau_{j+}(\tau ,
{\vec \eta}_i(\tau )),\vec{\eta}_{i}(\tau ))+\nonumber\\
&-&\int_{S_{\infty}}d\Sigma\:\vec{n}\cdot[
\check{\vec{E}}_{\perp IN}\times\check{\vec{B}}_{IN}+\nonumber\\
&+&\sum_{i=1}^NQ_{i}(\check{\vec{E}}_{\perp IN}\times\vec{\tilde{B}}_{(i+)}
(\tau_{i+},\vec \sigma )+\vec{\tilde{E}}_{\perp (i+)}(\tau_{i+},\vec \sigma )
\times\check{\vec{B}}_{IN})+\nonumber\\ 
&+&\sum_{i\neq j}Q_{i}Q_{j}
\vec{\tilde{E}}_{\perp (i+)}(\tau_{i+},\vec \sigma )\times\vec{\tilde{B}}
_{(j+)}(\tau_{j+},\vec \sigma )](\tau,\vec{\sigma})\, {\buildrel \circ \over
=}\, 0,
\label{f37}
\end{eqnarray}

\noindent where we used the formula [V is a sphere in the Wigner 
hyperplane and $S=\partial V$ its boundary with outer normal $\vec n$;
in Eq.(\ref{f37}), $S_{\infty}$ is the limit of S when the radius of the 
sphere goes to infinity]

\begin{equation}
\frac{dE}{d\tau}=
\frac{d}{d\tau}\int_{V}d^{3}\sigma\:
[\frac{\vec{E}^{2}_{\perp}+\vec{B}^{2}}{2}](\tau,\vec{\sigma})=
-\int_{S}d\Sigma\:\vec{n}\cdot(\vec{E}_{\perp}\wedge\vec{B})
(\tau,\vec{\sigma}).
\label{f38}
\end{equation}

In the nonrelativistic limit $|{\dot {\vec \eta}}_i(\tau )| < < 1$ and in wave 
zone [$\tau_{i+}(\tau ,\vec \sigma )\rightarrow \tau$, $\rho_{i+}(\tau_{i+}
(\tau ,\vec \sigma ),\vec \sigma )\rightarrow r(\tau ,\sigma )\approx |\vec
\sigma |\rightarrow \infty$] with ${\check {\vec A}}_{\perp IN}(\tau ,\vec
\sigma )=0$, the asymptotic limit of the retarded fields is

\begin{eqnarray}
{\check E}^{r}_{\perp RET,AS}(\tau ,\vec \sigma )&\approx&-P^{rs}_{\perp}(\vec
\sigma )\sum_{i=1}^N{{Q_i}\over {4\pi}} {{{\ddot \eta}^s_i(\tau )}\over
{|\vec \sigma |}},\nonumber \\ 
{\check B}^{r}_{RET,AS}(\tau ,\vec \sigma )&\approx&-P^{rs}_{\perp}(\vec \sigma
)\sum_{i=1}^N {{Q_i}\over {4\pi}} {{[\vec \sigma \times {\ddot {\vec \eta}}
_i(\tau )]^s}\over {|\vec \sigma |}},
\label{f39}
\end{eqnarray}

\noindent so that the ``Larmor formula" for the radiated energy become
[$\vec n={\hat {\vec r}}=\vec \sigma /|\vec \sigma |$]

\begin{eqnarray}
\frac{dE}{d\tau}&\approx&
\int_{S}d\Sigma\:\vec{n}\cdot(\check{\vec{E}}_{\perp RET,AS}\times 
\check{\vec{B}}_{RET,AS})(\tau ,\vec \sigma )=\nonumber\\
&=&\sum_{i\not= j}{{Q_iQ_j}\over {(4\pi )^2}}\int d\Omega
\:\vec{n}\cdot ({\ddot {\vec \eta}}_i(\tau )\times[\vec{n}\times
\ddot{\vec{\eta}}_j(\tau )])=\nonumber \\
&=&\sum_{i\not= j} {{Q_iQ_j}\over {(4\pi )^2}} \int d\Omega (\vec n\times 
{\ddot {\vec \eta}}_i(\tau ))\cdot (\vec n\times {\ddot {\vec \eta}}_j(\tau ))=
\nonumber \\
&=&\frac{2}{3}\sum_{i\not= j} {{Q_iQ_j}\over {(4\pi )^2}} {\ddot {\vec \eta}}_i
(\tau )\cdot {\ddot {\vec \eta}}_j(\tau ).
\label{f40}
\end{eqnarray}

The usual terms ${{Q^2_i}\over {(4\pi )^2}} {2\over 3} {\ddot {\vec \eta}}_i^2
(\tau )$ are absent due to the pseudoclassical conditions $Q^2_i=0$. Therefore,
at the pseudoclassical level, there is no radiation coming from single charges,
but only  interference radiation due to terms $Q_iQ_j$ with $i/not= j$. Since 
it is not possible to control whether the source is a single elementary charged
particle (only macroscopic sources are testable), this result is in accord
with macroscopic experimental facts.

For a single particle, N=1, the pseudoclassical equations (\ref{f36}) and
(\ref{f37}) become

\begin{eqnarray}
\frac{d}{d\tau}(\eta m \frac{\dot{\vec{\eta}}(\tau)}{\sqrt
{1-\dot{\vec{\eta}}^2(\tau) } })\, &{\buildrel \circ \over =}&\, Q[
\check{\vec{E}}_{\perp IN}(\tau,\vec{\eta}(\tau ))+\dot{\vec{\eta}}(\tau)\times
\check{\vec{B}}_{IN}(\tau,\vec{\eta})],\nonumber \\
&&{}\nonumber \\
&&\eta m\frac{\dot{\vec \eta}(\tau)}{\sqrt{1-\dot{\vec{\eta}}(\tau)}}
+Q\check{\vec{A}}_{\perp IN}(\tau,\vec{\eta}(\tau ))+\nonumber\\
&&+\int d^{3}\sigma
[(\check{\vec{E}}_{\perp IN}\times\check{\vec{B}}_{IN})(\tau,\vec{\sigma})+
Q(\check{\vec{E}}_{\perp IN}(\tau,\vec{\sigma})
\times\vec{\tilde{B}}_{(+)}(\tau_{+}(\tau,\vec{\sigma}),\vec \sigma )+
\nonumber \\
&+&\vec{\tilde{E}}_{\perp (+)}(\tau_{+}(\tau,\vec{\sigma}),\vec \sigma )\times
\check{\vec{B}}_{IN}(\tau,\vec{\sigma}))]
\, {\buildrel \circ \over =}\, 0,\nonumber \\
&&{}\nonumber \\
E_{rel}&=&\frac{\eta m}{\sqrt{1-\dot{\vec{\eta}}(\tau)^{2}}}+
\int d^{3}\sigma\:[\frac{\check{\vec{E}}^{2}_{\perp IN}+\check{\vec{B}}^{2}
_{IN}}{2}+\nonumber\\
&+&Q(\check{\vec{E}}_{\perp IN}\cdot\vec{\tilde{E}}_{\perp (+)}(\tau_{+},\vec
\sigma )+\check{\vec{B}}_{IN}\cdot\vec{\tilde{B}}_{(+)}(\tau_{+},\vec \sigma ))
](\tau,\vec{\sigma})=const.
\label{f41}
\end{eqnarray}

\begin{eqnarray}
\frac{d}{d\tau}\frac{\eta m}{\sqrt{1-\dot{\vec{\eta}}^{2}(\tau)}}\, &{\buildrel
\circ \over =}&\, Q\dot{\vec{\eta}}(\tau)\cdot\check{\vec{E}}_{\perp IN}(\tau,
\vec{\eta}(\tau ))+\int_{S_{as}}d\Sigma\:\vec{n}\cdot[
\check{\vec{E}}_{\perp IN}\times\check{\vec{B}}_{IN}+\nonumber\\
&+&Q(\check{\vec{E}}_{\perp IN}\times\vec{\tilde{B}}_{(+)}(\tau_{+},\vec 
\sigma )+\vec{\tilde{E}}_{\perp (+)}(\tau_{+},\vec \sigma )\times\check{\vec{B}
}_{IN})](\tau,\vec{\sigma}).
\label{f42}
\end{eqnarray}

The first of Eqs.(\ref{f41}) replaces the Abraham-Lorentz-Dirac equation 
[see for instance Ref.\cite{itz}] for
an electron in an external electromagnetic field [$Q=e\theta^{*}\theta$]

\begin{equation}
\frac{d}{d\tau}mu^{\mu}=eF^{\mu\nu}_{IN}(x)u_{\nu}+\frac{2}{3}\frac{e^{2}}{4\pi}
(\ddot{u}^{\mu}-
(u\cdot\ddot{u})u^{\mu}),\:\:u^{\mu}=\frac{\dot{x}^{\mu}}{\sqrt{\dot{x}^{2}}}.
\label{f43}
\end{equation}

The $e^2$ term contains: i) the term $(u\cdot\ddot{u})u^{\mu}=-\dot{u}^{2}u
^{\mu}$ associated with the Larmor emission of radiation; ii) the Schott
term $\ddot{u}^{\mu}$ producing violations of Einstein causality (either
runaway solutions or pre-acceleration). They are inseparable because, given
the Larmor term, the requirement of manifest Lorentz covariance forces the
appearance of the Schott term. The rest-frame instant form of dynamics has
manifest Wigner covariance, avoids the covariance problems in the
simultaneous description of particles and fields and has no term of order
$Q^2$ (only terms $Q_iQ_j$ with $i\not= j$) at the pseudoclassical level of
description of the electric charge.

Let us now come back to the general case with $\vec \lambda (\tau )\not= 0$,
for which we have  the Lagrangian (\ref{f16}). If we put $\vec \lambda (\tau )
={d\over {d\tau}} \vec g(\tau )$, then Eqs.(\ref{f16})-(\ref{f20}) become

\begin{eqnarray}
L_R(\tau)&=&-\sum_{i=1}^N\eta_{i}m_i\sqrt{1-
(\dot{\vec{\eta}}_{i}(\tau)+\dot{\vec{g}}(\tau))^{2}}+
\frac{1}{2}\sum_{i\neq j}\frac{Q_{i}Q_{j}}
{4\pi\mid\vec{\eta}_{i}(\tau )-\vec{\eta}_{j}(\tau )\mid}+\nonumber\\
&+&\int d^3\sigma \sum_{i=1}^N\delta^3(\vec \sigma -{\vec \eta}_i(\tau ))
Q_{i}[\dot{\vec{\eta}}_{i}(\tau)+\dot{\vec{g}}(\tau)]\cdot\check{\vec{A}}
_{\perp}(\tau,\vec \sigma)+\nonumber\\
&+&\int d^{3}\sigma {1\over 2}[\, ([{{\partial}\over {\partial \tau}}+{{d\vec
g(\tau )}\over {d\tau}}\cdot {{\partial}\over {\partial \vec \sigma}}]{\check
{\vec A}}_{\perp})^2-{\check {\vec B}}^2](\tau ,\vec \sigma ),\nonumber \\
&&{}\nonumber \\
&&\frac{d}{d\tau}[\eta_{i}m_{i}\frac{\dot{\vec{\eta}}_{i}(\tau)
+\dot{\vec{g}}(\tau)}{\sqrt{1-(\dot{\vec{\eta}}_{i}(\tau)
+\dot{\vec{g}}(\tau))^{2}}}+Q_{i}\check{\vec{A}}_{\perp}(\tau,\vec{\eta}_{i}
(\tau ))]\, {\buildrel \circ \over =}\nonumber\\
&{\buildrel \circ \over =}\, &-\sum_{k\neq i}
\frac{Q_{i}Q_{k}[{\vec \eta}_i(\tau )-{\vec \eta}_k(\tau )]}
{4\pi\mid\vec{\eta}_{i}(\tau )-\vec{\eta}_{k}(\tau )\mid^{3}}
+Q_i\sum_u[\dot{\eta}^{u}_{i}(\tau )+{\dot g}^u(\tau )]
{{\partial}\over {{\vec \eta}_{i}}}{\check A}^{u}_{\perp}(\tau,\vec{\eta}_{i}
(\tau )),\nonumber \\
&&{}\nonumber \\
&&[({{\partial}\over {\partial \tau}}+{{d\vec
g(\tau )}\over {d\tau}}\cdot {{\partial}\over {\partial \vec \sigma}})^2+
\triangle ]{\check A}^u_{\perp}(\tau ,\vec \sigma )\, {\buildrel \circ \over =}
\nonumber \\
&{\buildrel \circ \over =}&\, 
\sum_{i=1}^NQ_iP_{\perp}^{rs}(\vec \sigma )[\dot{\eta}^{s}_{i}(\tau )+
{\dot g}^s(\tau )]\delta^3(\vec{\sigma}-\vec{\eta}_{i}(\tau )),\nonumber \\
&&{}\nonumber \\
&&\sum_{i=1}^N[\eta_{i}m_i\frac{\dot{\vec{\eta}}_{i}(\tau)+\dot{\vec{g}}(\tau)}{
\sqrt{1-(\dot{\vec{\eta}}(\tau )+\dot{\vec{g}}(\tau))^{2}}}
+Q_{i}\check{\vec{A}}_{\perp}(\tau,\vec{\eta}_{i}(\tau ))]+\nonumber\\
&&+\int d^{3}\sigma\:
\sum_{r}[{{\partial {\check A}^r_{\perp}}\over {\partial \vec \sigma}}\cdot
({{\partial}\over {\partial \tau}}+{{d\vec
g(\tau )}\over {d\tau}}\cdot {{\partial}\over {\partial \vec \sigma}}){\check
A}^r_{\perp}](\tau,\vec{\sigma})\, {\buildrel \circ \over =}\, 0,\nonumber \\
&&{}\nonumber \\
E_{rel}&=&\sum_{i=1}^N\frac{\eta_{i}m_{i}}{\sqrt{1-(\dot{\vec{\eta}}_{i}(\tau)+
\dot{\vec{g}}(\tau))^{2}}}+\sum_{i>j}\frac{Q_{i}Q_{j}}
{4\pi\mid\vec{\eta}_{i}(\tau )-\vec{\eta}_{j}(\tau )\mid}+\nonumber \\
&+&\int d^{3}\sigma\: {1\over 2}[([{{\partial}\over {\partial \tau}}+{{d\vec
g(\tau )}\over {d\tau}}\cdot {{\partial}\over {\partial \vec \sigma}}]{\check
{\vec A}}_{\perp})^2(\tau,\vec{\sigma})+\check{\vec{B}}^{2}
(\tau,\vec{\sigma})]=const.
\label{ff1}
\end{eqnarray}

If we define the transformation $\tau =\tau^{'}$, $\vec \sigma ={\vec \sigma}
^{'}+\vec g(\tau^{'})$, we get ${{\partial}\over {\partial \tau^{'}}}=
{{\partial}\over {\partial \tau}}+{{d\vec
g(\tau )}\over {d\tau}}\cdot {{\partial}\over {\partial \vec \sigma}}$, 
${{\partial}\over {\partial {\vec \sigma}^{'}}}={{\partial}\over {\partial
\vec \sigma}}$ and the equation of motion for the transverse vector
potential and its solution become

\begin{eqnarray}
\Box^{'}{\check A}^r_{\perp}(\tau^{'},{\vec \sigma}^{'}+\vec g(\tau^{'}))\,
&{\buildrel \circ \over =}&\, \sum_{i=1}^NQ_iP^{rs}_{\perp}({\vec \sigma}^{'})
[{\dot \eta}^s_i(\tau^{'})+{\dot g}^s(\tau^{'})]\delta^3({\vec \sigma}^{'}+
\vec g(\tau^{'})-{\vec \eta}_i(\tau^{'})),\nonumber \\
&&{}\nonumber \\
{\check A}^r_{\perp RET}(\tau^{'},{\vec \sigma}^{'}+\vec g(\tau^{'}))&=&
{\check A}^r_{\perp IN}(\tau^{'},{\vec \sigma}^{'}+\vec g(\tau^{'}))+
\nonumber \\
&+&\sum_{i=1}^N{{Q_i}\over {2\pi}}P^{rs}_{\perp}({\vec \sigma}^{'})\int
d\bar \tau d^3\bar \sigma \theta (\tau^{'}-\bar \tau )\delta [(\tau^{'}-
\bar \tau )^2-({\vec \sigma}^{'}-{\vec {\bar \sigma}})^2]\nonumber \\
&\cdot& [{\dot \eta}^s_i
(\bar \tau )+{\dot g}^s(\bar \tau )]\delta^3({\vec {\bar \sigma}}+\vec g
(\bar \tau )-{\vec \eta}_i(\bar \tau ))=\nonumber \\
&=&{\check A}^r_{\perp IN}(\tau^{'},{\vec \sigma}^{'}+\vec g(\tau^{'}))+
\sum_{i=1}^N{{Q_i}\over {2\pi}}P^{rs}_{\perp}({\vec \sigma}^{'})\int d\bar \tau
\theta (\tau^{'}-\bar \tau )\nonumber \\
&&\delta [(\tau^{'}-\bar \tau )^2-({\vec \sigma}^{'}
+\vec g(\bar \tau )-{\vec \eta}_i(\bar \tau ))^2][{\dot \eta}^s_i(\bar \tau )+
{\dot g}^s(\bar \tau )].
\label{ff2}
\end{eqnarray}

If $\tau^{'}_{i+}(\tau^{'},{\vec \sigma}^{'}+\vec g(\tau^{'}))$ is the retarded
solution of the delta-function and we define the functions ${\vec r}^{'}_{i+}
(\tau^{'}_{i+}(\tau^{'},{\vec \sigma}^{'}+\vec g(\tau^{'})),{\vec \sigma}^{'}+
\vec g(\tau^{'}))={\vec \sigma}^{'}+\vec g(\tau^{'})-{\vec \eta}_i(\tau^{'}
_{i+}(\tau^{'},{\vec \sigma}^{'}+\vec g(\tau^{'})))$ and the associated
$\rho^{'}_{i+}$, we get

\begin{eqnarray}
{\check A}^r_{\perp RET}(\tau^{'},{\vec \sigma}^{'}+\vec g(\tau^{'}))&=&
{\check A}^r_{\perp IN}(\tau^{'},{\vec \sigma}^{'}+\vec g(\tau^{'}))+
\nonumber \\
&+&\sum_{i=1}^N{{Q_i}\over {4\pi}}P^{rs}_{\perp}({\vec \sigma}^{'})
{{ {\dot \eta}^s_i(\tau^{'}_{i+}(\tau^{'},{\vec \sigma}^{'}+\vec g(\tau^{'})))+
{\dot g}^s(\tau^{'}_{i+}(\tau^{'},{\vec \sigma}^{'}+\vec g(\tau^{'})))}\over
{ \rho^{'}_{i+}(\tau^{'}_{i+}(\tau^{'},{\vec \sigma}^{'}+\vec g(\tau^{'})),
{\vec \sigma}^{'}+\vec g(\tau^{'}))}}.
\label{ff3}
\end{eqnarray}

Therefore, one could recover all the previous results with $\vec \lambda (\tau )
={\dot {\vec g}}(\tau )\not= 0$.

If we put Eq.(\ref{ff3}) with ${\check {\vec A}}_{\perp IN}=0$ in the
Lagrangian of Eq.(\ref{ff1}), we get the effective Fokker action in the
rest-frame instant form [one should make a careful analysis of the boundary
terms in the variation of $L_R(\tau )$ following Ref.\cite{f1}, before
claiming the equivalence of this effective Fokker action to a subspace of
solutions of the original theory].

Following Ref.\cite{f4} and in particular Refs.\cite{g1,g2,g3}, one can 
substitute retarded particle coordinates and velocities with instantaneous
[in $\tau$] coordinates and accelerations of all orders. There are two
methods for doing this in Refs.\cite{g1,g2,g3}. It would be interesting
to check whether (with one of
these methods) one can confirm in a Wigner-covariant way the non-covariant
result of Gordeyev that starting with an instantaneous expansion in
accelerations of retarded equations one gets instantaneous actions for the
accelerations of the type ${1\over 2}(retarded+advanced)$, but with the advanced
part being a total $\tau$-derivative (so that it does not contribute to the
equations of motion). See also Ref.\cite{f2}.

Let us note Ref.\cite{f3} is the only attempt to study the Dirac constraints
originating from actions depending from accelerations of all orders [and, 
assuming they are equivalent to Fokker actions, also originating
from Fokker actions].

A connected open problem is the comparison for N=2 of the invariant mass
$\epsilon_s=H_{rel}$ of Eq.(\ref{f12}), when one puts into it the retarded
[or the ${1\over 2}(retarded+advanced)$] Lienard-Wiechert solutions (\ref{f31}),
(\ref{f34}), (\ref{f35}), with ${\check {\vec A}}_{\perp IN}(\tau ,\vec 
\sigma )=0$, with the sum of the pair of first class constraints with 
instantaneous potentials\cite{dv,tod,ll,crater,saz}, whose quantization gives
coupled Klein-Gordon equations for two spin zero particles [see Refs.
\cite{crater,saz,crater1,saz1} for similar equations for Dirac particles 
deriving from pairs of first class constraints for spinning particles]. These
models were generated as phenomenological approximations to the Bethe-Salpeter
equation, by reducing it in instantaneous approximations to a 3-dimensional
equation (with the elimination of the spurious abnormal sectors of relative
energy excitations) of the Lippmann-Schwinger type and then to the equation of
the quasipotential approach [sse the bibliography of the quoted references],
which Todorov\cite{tod} reformulated as a pair of first class constraints at 
the classical level. In Ref.\cite{saz} it is directly shown how the normal 
sectors of the Bethe-Salpeter equation are connected with the quantization of 
pairs of first class constraints with instantaneous (in general nonlocal, but
approximable with local) potentials like in Todorov's examples. Instead, in
Refs.\cite{crater} it is shown how to derive the Todorov potential for the
electromagnetic case from Tetrode-Fokker-Feynman-Wheeler electrodynamics with
scalar and vector potentials [this theory is connected with ${1\over 2}(retarded
+advanced)$ solutions with no incoming radiation (adjunct Lienard-Wiechert 
fields) of Maxwell equations with particle currents in the Lorentz gauge]; 
besides the Coulomb potential, at the order $1/c^2$ one gets the Darwin
potential (becoming the Breit one at the quantum level), which is known to be
phenomenologically correct. With only retarded Lienard-Wiechert potentials
Eq.(\ref{f12}) becomes

\begin{eqnarray}
\epsilon_s&=&H_{rel}=\eta_1\sqrt{m_1^2+[{\check {\vec \kappa}}_1(\tau )-Q_1Q_2
{\vec {\tilde A}}_{\perp (2+)}(\tau_{(2+)}(\tau,{\vec \eta}_1(\tau )),{\vec
\eta}_1(\tau ))]^2}+\nonumber \\
&+&\eta_2\sqrt{m_2^2+[{\check {\vec \kappa}}_2(\tau )-Q_1Q_2
{\vec {\tilde A}}_{\perp (1+)}(\tau_{(1+)}(\tau,{\vec \eta}_2(\tau )),{\vec
\eta}_2(\tau ))]^2}+{{Q_1Q_2}\over {4\pi |{\vec \eta}_1(\tau )-{\vec \eta}_2
(\tau )|}}+\nonumber \\
&+&Q_1Q_2 \int d^3\sigma [{\vec {\tilde E}}_{\perp (1+)}(\tau_{(1+)}(\tau ,
\vec \sigma ),\vec \sigma )\cdot {\vec {\tilde E}}_{\perp (2+)}(\tau_{(2+)}
(\tau ,\vec \sigma ),\vec \sigma )+\nonumber \\
&+&{\vec {\tilde B}}_{(1+)}(\tau_{(1+)}(\tau ,
\vec \sigma ),\vec \sigma )\cdot {\vec {\tilde B}}_{(2+)}(\tau_{(2+)}(\tau 
,\vec \sigma ),\vec \sigma ).
\label{z1}
\end{eqnarray}

\noindent How, also forgetting the last term, 
to reexpress it only in terms of particle coordinates and momenta
[this problem is connected with the previous one of the Hamiltonian formulation
of Fokker actions]? How to check with a $1/c^2$ expansion whether the Darwin 
potential is already present without using ${1\over 2}(retarded+advanced)$
solutions [this is connected with Gordeyev approach]?

Let us remark that, if in Eqs.(\ref{f17}), (\ref{f19}) and (\ref{f20}) one
adds the second class constraints 
$\check{\vec{A}}_{\perp}(\tau ,\vec \sigma )=\check{\vec{\pi}}_{\perp}
(\tau ,\vec \sigma )=0$ one selects the sector of phase space describing 
classical bound states without any kind of radiation
[namely one looks for solutions of the  charged
N-body problem with instantaneous Coulomb interaction but without radiation 
fields]. One gets the equations

\begin{eqnarray}
\frac{d}{d\tau}(\eta_{i}m_{i}\frac{\dot{\vec{\eta}}_{i}(\tau)}{\sqrt{
1-\dot{\vec{\eta}}_{i}^{2}(\tau)}})&=&
-\sum_{k\neq i}\frac{Q_{i}Q_{k}[{\vec \eta}_i(\tau )-{\vec \eta}_k(\tau )]}
{4\pi\mid\vec{\eta}_{i}(\tau )-\vec{\eta}_{k}(\tau )\mid^{3}}\nonumber\\
\sum_{i=1}^N\eta_{i}m_{i}\frac{\dot{\vec{\eta}}_{i}(\tau)}{\sqrt{
1-\dot{\vec{\eta}}_{i}^{2}(\tau)}}&=&0\nonumber\\
E_{rel}&=&\sum_{i}\frac{\eta_{i}m_{i}}{\sqrt{
1-\dot{\vec{\eta}}_{i}^{2}(\tau)}}+\sum_{i\neq k}\frac{Q_{i}Q_{k}}
{4\pi\mid\vec{\eta}_{i}(\tau )-\vec{\eta}_{k}(\tau )\mid}=const.
\label{f44}
\end{eqnarray}  

\noindent For N=2 and $m_{1}=m_{2}=m$ one knows the solutions \cite{gst}:
there is a rosetta motion without spiral fall on the center.

Finally, in the electromagnetic case, the distribution function on the charge
Grassmann variables\cite{casala} is

\begin{equation}
\rho(\theta_{i},\theta^{*}_{i})=
\prod_{i}(1+\theta^{*}_{i}\theta_{i}).
\label{f45}  
\end{equation}

\noindent It satisfies the positivity condition only on analytic functions
of the $\theta_i$'s

\begin{equation}
f(\theta_{i})=f_{0}+\sum_{i}f_{1i}\theta_{i}.
\label{f46}
\end{equation}

\noindent The classical theory is recovered by making the mean of Grassmann-
valued observables with this distribution function. For the electric charges
one has $< Q_i >=< e_i\theta^{*}_i\theta_i > =e_i$. As noted in 
Refs.\cite{casala}, the processes of taking the mean of the equations of motion
and then solving the classical equations [with the standard electromagnetic
divergences and causality pathologies] or of solving the pseudoclassical field 
equations and then taking the mean [no divergences and no causality problems]
do not commute. Since in the latter case there is a regularization of the
electromagnetic self-energy, it would be important to learn how to quantize
these solutions.

\vfill\eject

\section{Scalar Electrodynamics on Spacelike Hypersurfaces}

Let us consider the action describing a charged Klein Gordon field interacting
with the electromagnetic field on spacelike hypersurfaces following the scheme
of Ref.\cite{lus1}

\begin{eqnarray}
S&=& \int d\tau d^3\sigma N(\tau ,\vec \sigma )\sqrt{\gamma (\tau ,\vec \sigma )
}\nonumber \\
&&\{ g^{\tau\tau} (\partial_{\tau}+ieA_{\tau}) \phi^{*}\, (\partial_{\tau}-ieA
_{\tau}) \phi +\nonumber \\
&&+g^{\tau \check r} [(\partial_{\tau}+ieA_{\tau}) \phi^{*}\, (\partial
_{\check r}-ieA_{\check r}) \phi +(\partial_{\check r}+ieA_{\check r}) \phi^{*}\,
(\partial_{\tau}-ieA_{\tau}) \phi ] +\nonumber \\
&&+g^{\check r\check s}(\partial_{\check r}+ieA_{\check r}) \phi^{*}\, (\partial
_{\check s}-ieA_{\check s}) \phi - m^2 \phi^{*}\phi -{1\over 4}g^{\check A\check
C}g^{\check B\check D}F_{\check A\check B}F_{\check C\check D}\, \}
(\tau ,\vec \sigma )=\nonumber \\
&&=\int d\tau d^3\sigma \sqrt{\gamma (\tau ,\vec \sigma )} \{ {1\over N}
[\partial_{\tau}+ieA_{\tau} -N^{\check r}(\partial_{\check r}+ieA_{\check r})]
\phi^{*}\nonumber \\
&&[\partial_{\tau}-ieA_{\tau} -N^{\check s}(\partial_{\check s}-ieA_{\check s})]
\phi + N [\gamma^{\check r\check s}(\partial_{\check r}+ieA_{\check r})\phi^{*}
(\partial_{\check s}-ieA_{\check s})\phi -m^2\phi^{*}\phi ]-\nonumber \\
&&-{1\over {2N}}(F_{\tau\check r}-N^{\check u}F_{\check u\check r})\gamma
^{\check r\check s}(F_{\tau \check s}-N^{\check v}F_{\check v\check s})-{N\over
4} \gamma^{\check r\check s}\gamma^{\check u\check v}F_{\check r\check u}
F_{\check s\check v}\, \}(\tau ,\vec \sigma ).
\label{b1}
\end{eqnarray}

\noindent where the configuration variables are $z^{\mu}(\tau ,\vec \sigma )$,
$\phi (\tau ,\vec \sigma )={\tilde \phi}(z(\tau ,\vec \sigma ))$ and $A_{\check 
A}(\tau ,\vec \sigma )=z^{\mu}_{\check A}(\tau ,\vec \sigma )A_{\mu}(z
(\tau ,\vec \sigma ))$ [$\tilde \phi (z)$ and $A_{\mu}(z)$ are the standard
Klein-Gordon field and electromagnetic potential, which do not know the
embedding of the spacelike hypersurface $\Sigma$ in Mikowski spacetime like
$\phi$ and $A_{\check A}$].

Since $z^{\mu}_{\tau}=Nl^{\mu}+N^{\check r}z^{\mu}_{\check r}$, one has 
${{\partial}\over {\partial z^{\mu}_{\tau}}}=l_{\mu}{{\partial}\over {\partial
N}}+z_{\check s\mu}\gamma^{\check s\check r}{{\partial}\over {\partial N^{\check
r}}}$. Therefore, the canonical momenta are

\begin{eqnarray}
\pi^{\tau}(\tau ,\vec \sigma )&=&{{\partial L}\over {\partial \partial_{\tau}
A_{\tau}(\tau ,\vec \sigma )}}=0,\nonumber \\
\pi^{\check r}(\tau ,\vec \sigma )&=&{{\partial L}\over {\partial \partial
_{\tau}A_{\check r}(\tau ,\vec \sigma )}}=\nonumber \\
&=&-{{\sqrt{\gamma}(\tau ,\vec \sigma )}\over
{N(\tau ,\vec \sigma )}}\gamma^{\check r\check s}(\tau ,\vec \sigma )(F_{\tau
\check s}-N^{\check u}F_{\check u\check s})(\tau ,\vec \sigma ),\nonumber \\
\pi_{\phi}(\tau ,\vec \sigma )&=&{{\partial L}\over {\partial \partial_{\tau}
\phi (\tau ,\vec \sigma )}}=\nonumber \\
&=&{{\sqrt{\gamma}(\tau ,\vec \sigma )}\over
{N(\tau ,\vec \sigma )}}[\partial_{\tau}+ieA_{\tau}-N^{\check r}(\partial
_{\check r}+ieA_{\check r})](\tau ,\vec \sigma ) \phi^{*}(\tau ,\vec \sigma ),
\nonumber \\
\pi_{\phi^{*}}(\tau ,\vec \sigma )&=&{{\partial L}\over {\partial \partial
_{\tau}\phi^{*}(\tau ,\vec \sigma )}}=\nonumber \\
&=&{{\sqrt{\gamma}(\tau ,\vec \sigma )}\over
{N(\tau ,\vec \sigma )}} [\partial_{\tau}-ieA_{\tau}-N^{\check r}(\partial
_{\check r}-ieA_{\check r})](\tau ,\vec \sigma ) \phi (\tau ,\vec \sigma ),
\nonumber \\
\rho_{\mu}(\tau ,\vec \sigma )&=&-{{\partial L}\over {\partial \partial_{\tau}
z^{\mu}(\tau ,\vec \sigma )}}=\nonumber \\
&=&l_{\mu}(\tau ,\vec \sigma ) \{ {{\pi_{\phi}
\pi_{\phi^{*}}}\over {\sqrt{\gamma}}}-\sqrt{\gamma} [\gamma^{\check r\check s}
(\partial_{\check r}+ieA_{\check r})\phi^{*} (\partial_{\check s}-ieA_{\check 
s})\phi -\nonumber \\
&&-m^2\phi^{*}\phi] +{1\over {2\sqrt{\gamma}}}\pi^{\check r}g_{\check r\check 
s}\pi^{\check s}-{{\sqrt{\gamma}}\over 4}\gamma^{\check r\check s}\gamma
^{\check u\check v}F_{\check r\check u}F_{\check s\check v} \}(\tau ,\vec 
\sigma )+\nonumber \\
&&+z_{\check s\mu}(\tau ,\vec \sigma )\gamma^{\check r\check s}(\tau ,\vec 
\sigma ) \{ \pi_{\phi^{*}}(\partial_{\check r}+ieA_{\check r})\phi^{*}+
\pi_{\phi} (\partial_{\check r}-ieA_{\check r})\phi-F_{\check r\check u}
\pi^{\check u} \} (\tau ,\vec \sigma ).
\label{b2}
\end{eqnarray}

Therefore, one has the following primary constraints

\begin{eqnarray}
\pi^{\tau}(\tau ,\vec \sigma )&\approx& 0,\nonumber \\
{\cal H}_{\mu}(\tau ,\vec \sigma )&=& \rho_{\mu}(\tau ,\vec \sigma )-
\nonumber \\
&&-l_{\mu}(\tau ,\vec \sigma ) \{ {{\pi_{\phi}
\pi_{\phi^{*}}}\over {\sqrt{\gamma}}}-\sqrt{\gamma} [\gamma^{\check r\check s}
(\partial_{\check r}+ieA_{\check r})\phi^{*} (\partial_{\check s}-ieA_{\check 
s})\phi -\nonumber \\
&&-m^2\phi^{*}\phi] +{1\over {2\sqrt{\gamma}}}\pi^{\check r}g_{\check r\check 
s}\pi^{\check s}-{{\sqrt{\gamma}}\over 4}\gamma^{\check r\check s}\gamma
^{\check u\check v}F_{\check r\check u}F_{\check s\check v} \}(\tau ,\vec 
\sigma )+\nonumber \\
&&+z_{\check s\mu}(\tau ,\vec \sigma )\gamma^{\check r\check s}(\tau ,\vec 
\sigma ) \{ \pi_{\phi^{*}}(\partial_{\check r}+ieA_{\check r})\phi^{*}+
\pi_{\phi} (\partial_{\check r}-ieA_{\check r})\phi-F_{\check r\check u}
\pi^{\check u} \} (\tau ,\vec \sigma ) \approx 0,
\label{b3}
\end{eqnarray}

\noindent and the following Dirac Hamiltonian [$\lambda (\tau ,\vec \sigma )$
and $\lambda^{\mu}(\tau ,\vec \sigma )$ are Dirac multiplier]

\begin{equation}
H_D=\int d^3\sigma [-A_{\tau}(\tau ,\vec \sigma ) \Gamma (\tau ,\vec \sigma )+
\lambda (\tau ,\vec \sigma )\pi^{\tau}(\tau ,\vec \sigma )+\lambda^{\mu}
(\tau ,\vec \sigma ){\cal H}_{\mu}(\tau ,\vec \sigma )].
\label{b4}
\end{equation}

By using the Poisson brackets

\begin{eqnarray}
\{ z^{\mu}(\tau ,\vec \sigma ),\rho_{\nu}(\tau ,{\vec \sigma}^{'}) \} &=&\eta
^{\mu}_{\nu} \delta^3(\vec \sigma -{\vec \sigma}^{'}),\nonumber \\
\{ A_{\check A}(\tau ,\vec \sigma ),\pi^{\check B}(\tau ,{\vec \sigma}^{'}) \}
&=&\eta^{\check B}_{\check A} \delta^3(\vec \sigma -{\vec \sigma}^{'}),
\nonumber \\
\{ \phi (\tau ,\vec \sigma ),\pi_{\phi}(\tau ,{\vec \sigma}^{'}) \} &=&
\{ \phi^{*}(\tau ,\vec \sigma ),\pi_{\phi^{*}}(\tau ,{\vec \sigma}^{'}) \} =
\delta^3(\vec \sigma -{\vec \sigma}^{'}),
\label{b5}
\end{eqnarray}

\noindent one finds that
the time constancy of the primary constraints implies the existence of only 
one secondary constraint

\begin{equation}
\Gamma (\tau ,\vec \sigma ) =\partial_{\check r}\pi^{\check r}(\tau ,\vec 
\sigma )+ie (\pi_{\phi^{*}} \phi^{*}-\pi_{\phi} \phi 0(\tau ,\vec \sigma )
\approx 0.
\label{b6}
\end{equation}

One can verify that these constraints are first class with the algebra given
in Eqs.(125) of Ref.\cite{lus1}.

The Poincare' generators are like in Eq.(\ref{a28}).

Following Ref.\cite{lus1} (see also Sections I, II of this paper), 
we can restrict ourselves to spacelike hyperplanes
$z^{\mu}(\tau ,\vec \sigma )=x_s^{\mu}(\tau )+b^{\mu}_{\check r}
\sigma^{\check r}$
where the normal $l^{\mu}=\epsilon^{\mu}_{\alpha\beta\gamma}b_1^{\alpha}(\tau )
b_2^{\beta}(\tau )b_3^{\gamma}(\tau )$ is $\tau$-independent. Using the results
of that paper one finds that $J^{\mu\nu}_s=x_s^{\mu}p_s^{\nu}-x_s^{\nu}p_s
^{\mu}+S^{\mu\nu}_s$ and that the constraints are reduced to the following ones

\begin{eqnarray}
\pi^{\tau}(\tau ,\vec \sigma )&\approx& 0,\nonumber \\
\Gamma (\tau ,\vec \sigma )&=&-\vec \partial \vec \pi (\tau ,\vec \sigma )+
ie [\pi_{\phi^{*}}\phi^{*}-\pi_{\phi}\phi ](\tau ,\vec \sigma )\approx 0,
\nonumber \\
{\tilde {\cal H}}^{\mu}(\tau )&=& \int d^3\sigma {\cal H}^{\mu}(\tau ,\vec 
\sigma )=\nonumber \\
&=&p^{\mu}_s-l^{\mu} \{ {1\over 2} \int d^3\sigma [{\vec \pi}^2+{\vec B}^2]
(\tau ,\vec \sigma )+\nonumber \\
&+&\int d^3\sigma [\pi_{\phi^{*}}\phi^{*} \pi_{\phi}+(\vec \partial +ie\vec A)
\phi^{*} \cdot (\vec \partial -ie\vec A)\phi +m^2\phi^{*}\phi ](\tau ,\vec 
\sigma ) \}-\nonumber \\
&-&b^{\mu}_{\check r}(\tau ) \{ \int d^3\sigma (\vec \pi \times \vec B)_{\check 
r}(\tau ,\vec \sigma )+\int d^3\sigma [\pi_{\phi^{*}}(\partial_{\check r}+
ieA_{\check r})\phi^{*}+\nonumber \\
&+&\pi_{\phi} (\partial_{\check r}-ieA_{\check r}) \phi ](\tau ,\vec \sigma ) \}
\approx 0,\nonumber \\
{\tilde {\cal H}}^{\mu\nu}(\tau )&=&b^{\mu}_{\check r}(\tau ) \int d^3\sigma
\sigma^{\check r} {\cal H}^{\nu}(\tau ,\vec \sigma )-b^{\nu}_{\check r}(\tau )
\int d^3\sigma \sigma^{\check r} {\cal H}^{\mu}(\tau ,\vec \sigma )=\nonumber \\
&=&S^{\mu\nu}_s-(b^{\mu}_{\check r}(\tau )l^{\nu}-b^{\nu}_{\check r}(\tau )
l^{\mu}) [{1\over 2}\int d^3\sigma \sigma^{\check r} ({\vec \pi}^2+{\vec B}^2)
(\tau ,\vec \sigma )+\nonumber \\
&+&\int d^3\sigma \sigma^{\check r}[\pi_{\phi^{*}}\pi_{\phi}+(\vec \partial +
ie\vec A)\phi^{*}\cdot (\vec \partial -ie\vec A)\phi +m^2\phi^{*}\phi ](\tau ,
\vec \sigma ) \}+\nonumber \\
&+&(b^{\mu}_{\check r}(\tau )b^{\nu}_{\check s}(\tau )-b^{\nu}_{\check r}(\tau 
)b^{\mu}_{\check s}(\tau )) \{ \int d^3\sigma \sigma^{\check r} (\vec \pi 
\times \vec B)_{\check s}(\tau ,\vec \sigma )+\nonumber \\
&+&\int d^3\sigma \sigma^{\check r} [\pi_{\phi^{*}}(\partial_{\check s}+ieA
_{\check s})\phi^{*}+\pi_{\phi} (\partial_{\check s}-ieA_{\check s})\phi ]
(\tau ,\vec \sigma ) \} \approx 0.
\label{b8}
\end{eqnarray}

The configuration variables are reduced from $z^{\mu}(\tau ,\vec \sigma )$,
$A_{\check A}(\tau ,\vec \sigma )$, $\phi (\tau ,\vec \sigma )$, $\phi^{*}
(\tau ,\vec \sigma )$ to $x_s^{\mu}(\tau )$, to the six independent degrees of
freedom hidden in the orthonormal tetrad $b^{\mu}_{\check A}$ [$b^{\mu}_{\tau}=
l^{\mu}$], $A_{\check A}(\tau ,\vec \sigma )$, $\phi (\tau ,\vec \sigma )$, 
$\phi^{*}(\tau ,\vec \sigma )$, with the associated momenta [six degrees of 
freedom hidden in $S^{\mu\nu}_s$ are the momenta conjugate to those hidden in
the tetrad; see Ref.\cite{lus1} for the associated Dirac brackets].

If one selects all the configurations of the system with timelike total momentum
[$p^2_s > 0$], one can restrict oneself to the special Wigner hyperplanes
orthogonal to $p^{\mu}_s$ itself. The effect of this gauge fixing is a
canonical reduction to a phase space spanned only by the variables ${\tilde x}
^{\mu}_s(\tau )$, $p^{\mu}_s$, $A_{\tau}(\tau ,\vec \sigma )$, $\pi^{\tau}
(\tau ,\vec \sigma )$, $\vec A(\tau ,\vec \sigma )$, $\vec \pi (\tau ,\vec 
\sigma )$, $\phi (\tau ,\vec \sigma )$, $\pi_{\phi}(\tau ,\vec \sigma )$,
$\phi^{*}(\tau ,\vec \sigma )$, $\pi_{\phi^{*}}(\tau ,\vec \sigma )$, with
standard Dirac brackets. 

The only surviving constraints are [$\epsilon_s=\eta_s\sqrt{p^2_s}$]

\begin{eqnarray}
\pi^{\tau}(\tau ,\vec \sigma )&\approx& 0,\nonumber \\
\Gamma (\tau ,\vec \sigma )&=&-\vec \partial \vec \pi (\tau ,\vec \sigma )+
ie [\pi_{\phi^{*}}\phi^{*}-\pi_{\phi}\phi ](\tau ,\vec \sigma )\approx 0,
\nonumber \\
{\cal H}(\tau )&=& \epsilon_s-\{ {1\over 2} \int d^3\sigma ({\vec \pi}^2+{\vec 
B}^2)(\tau ,\vec \sigma )+\nonumber \\
&+&\int d^3\sigma [\pi_{\phi^{*}}\phi^{*} \pi_{\phi}+(\vec \partial +ie\vec A)
\phi^{*} \cdot (\vec \partial -ie\vec A)\phi +m^2\phi^{*}\phi ](\tau ,\vec 
\sigma ) \} \approx 0,\nonumber \\
{\vec {\cal H}}_p(\tau )&=& \int d^3\sigma (\vec \pi \times \vec B)(\tau ,\vec 
\sigma )+\nonumber \\
&+&\int d^3\sigma [\pi_{\phi^{*}} (\vec \partial +ie\vec A)\phi^{*} +\pi_{\phi}
(\vec \partial -ie\vec A)\phi ](\tau ,\vec \sigma ) \approx 0.
\label{b9}
\end{eqnarray}

Always following Ref.\cite{lus1}, it can be shown that the Lorentz generators
take the following form

\begin{eqnarray}
J^{ij}_s&=&{\tilde x}_s^ip_s^j-{\tilde x}_s^jp_s^i+\delta^{ir}\delta^{js}
{\bar S}_s^{rs},\nonumber \\
J^{oi}_s&=&{\tilde x}_s^op_s^i-{\tilde x}^i_sp^o_s-{{\delta^{ir}{\bar S}_s^{rs}
p_s^s}\over {p^o_s+\epsilon_s}},\nonumber \\
&&{}\nonumber \\
{\bar S}_s^{rs}&=&\int d^3\sigma \{ \sigma^r (\vec \pi \times \vec B)^s(\tau 
,\vec \sigma )-\sigma^s (\vec \pi \times \vec B)^r(\tau ,\vec \sigma ) \}+
\nonumber \\
&+&\int d^3\sigma \{ \sigma^r[\pi_{\phi^{*}}(\partial^s+ieA^s)\phi^{*}+\pi
_{\phi}(\partial^s-ieA^s)\phi](\tau ,\vec \sigma )-(r \leftrightarrow s) \}.
\label{b10}
\end{eqnarray}

To make the reduction to Dirac's observables with respect to the electromagnetic
gauge transformations, let us recall \cite{lusa,lv1} that the electromagnetic
gauge degrees of freedom are described by the two pairs of conjugate variables
$A_{\tau}(\tau ,\vec \sigma )$, $\pi_{\tau}(\tau ,\vec \sigma )[\approx 0]$,
$\eta_{em}(\tau ,\vec \sigma )=-{1\over {\triangle}} {{\partial}\over {\partial
\vec \sigma}}\cdot \vec A(\tau ,\vec \sigma )$, $\Gamma (\tau ,\vec \sigma )
[\approx 0]$, so that we have the decompositions

\begin{eqnarray}
A^r(\tau ,\vec \sigma )&=&{{\partial}\over {\partial \sigma^r}} \eta_{em}(\tau 
,\vec \sigma )+A^r_{\perp}(\tau ,\vec \sigma ),\nonumber \\
\pi^r(\tau ,\vec \sigma )&=&\pi^r_{\perp}(\tau ,\vec \sigma )+\nonumber \\
&+&{1\over {\triangle}} {{\partial}\over {\partial \sigma^r}} [-\Gamma (\tau 
,\vec \sigma )+ie(\pi_{\phi^{*}}\phi^{*}-\pi_{\phi}\phi )(\tau ,\vec \sigma )]
\approx 0,\nonumber \\
&&{}\nonumber \\
\lbrace A^r_{\perp}(\tau ,\vec \sigma )&,&\pi^s_{\perp}(\tau ,{\vec \sigma}^{'})
\rbrace
=-P^{rs}_{\perp}(\vec \sigma ) \delta^3(\vec \sigma -{\vec \sigma}^{'}),
\label{b11}
\end{eqnarray}

\noindent where $P^{rs}_{\perp}(\vec \sigma )=\delta^{rs}+{{\partial^r
\partial^s}\over {\triangle}}$, $\triangle =-{\vec \partial}^2$. Then, we have

\begin{eqnarray}
\int d^3\sigma &&{\vec \pi}^2(\tau ,\vec \sigma )= \int d^3\sigma {\vec \pi}^2
_{\perp}(\tau ,\vec \sigma )-\nonumber \\
&-&{{e^2}\over {4\pi}} \int d^3\sigma_1d^3\sigma_2 {{i(\pi_{\phi^{*}}\phi^{*}-
\pi_{\phi}\phi )(\tau ,{\vec \sigma}_1)\, i(\pi_{\phi^{*}}\phi^{*}-
\pi_{\phi}\phi )(\tau ,{\vec \sigma}_2)}\over {|{\vec \sigma}_1-{\vec \sigma}
_2|}}.
\label{b12}
\end{eqnarray}

Since we have

\begin{eqnarray}
\lbrace \phi (\tau ,\vec \sigma ),\Gamma (\tau ,{\vec \sigma}^{'})\rbrace &=&
ie \phi(\tau ,\vec \sigma ) \delta^3(\vec \sigma -{\vec \sigma}^{'}),
\nonumber \\
\lbrace \pi_{\phi}(\tau ,\vec \sigma ),\Gamma (\tau ,{\vec \sigma}^{'})\rbrace
&=&-ie \pi_{\phi}(\tau ,\vec \sigma ) \delta^3(\vec \sigma -{\vec \sigma}^{'}),
\label{b13}
\end{eqnarray}

\noindent the Dirac observables for the Klein-Gordon field are

\begin{eqnarray}
\hat \phi (\tau ,\vec \sigma )&=&[{\hat \phi}^{*}(\tau ,\vec \sigma )]^{*}=
e^{ie\eta_{em}(\tau ,\vec \sigma )} \phi (\tau ,\vec \sigma ),\nonumber \\
{\hat \pi}_{\phi}(\tau ,\vec \sigma )&=&[{\hat \pi}_{\phi^{*}}(\tau ,\vec 
\sigma )]^{*}=e^{-ie\eta_{em}(\tau ,\vec \sigma )} \pi_{\phi}(\tau ,\vec 
\sigma ),\nonumber \\
&&{}\nonumber \\
\lbrace \hat \phi (\tau ,\vec \sigma )&,&\Gamma (\tau ,{\vec \sigma}^{'})\rbrace
=\lbrace {\hat \pi}_{\phi}(\tau ,\vec \sigma ),\Gamma (\tau ,{\vec \sigma}^{'})
\lbrace =0.
\label{b14}
\end{eqnarray}

The constraints take the following form

\begin{eqnarray}
{\cal H}(\tau )&=&\epsilon_s- \{ \, {1\over 2}\int d^3\sigma ({\vec \pi}^2
_{\perp}+{\vec B}^2)(\tau ,\vec \sigma )+\nonumber \\
&+&\int d^3\sigma [{\hat \pi}_{\phi^{*}}{\hat \pi}_{\phi}+(\vec \partial +ie
{\vec A}_{\perp}){\hat \phi}^{*}\cdot (\vec \partial -ie{\vec A}_{\perp})\hat 
\phi +m^2{\hat \phi}^{*}\hat \phi ](\tau ,\vec \sigma )-\nonumber \\
&-&{{e^2}\over {8\pi}}\int d^3\sigma_1d^3\sigma_2 {{i({\hat \pi}_{\phi^{*}}{\hat
\phi}^{*}-{\hat \pi}_{\phi}\hat \phi )(\tau ,{\vec \sigma}_1)\, i({\hat \pi}
_{\phi^{*}}{\hat \phi}^{*}-{\hat \pi}_{\phi}\hat \phi )(\tau ,{\vec \sigma_2)}}
\over {|{\vec \sigma}_1-{\vec \sigma}_2|}} \} ,\nonumber \\
&&{}\nonumber \\
{\vec {\cal H}}_p(\tau )&=&\int d^3\sigma ({\vec \pi}_{\perp}\times \vec B)
(\tau ,\vec \sigma )+\int d^3\sigma ({\hat \pi}_{\phi^{*}}\vec \partial {\hat 
\phi}^{*}+{\hat \pi}_{\phi}\vec \partial \hat \phi )(\tau ,\vec \sigma ) 
\approx 0,
\label{b15}
\end{eqnarray}

\noindent where the Coulomb self-interaction appears in the invariant mass and
where the 3 constraints defining the rest frame do not depend on the interaction
since we are in an instant form of the dynamics. The final form of the 
rest-frame spin tensor is

\begin{equation}
{\bar S}^{rs}_s=\int d^3\sigma \{ \sigma^r [({\vec \pi}_{\perp}\times \vec B)^s
+{\hat \pi}_{\phi^{*}}\partial^s{\hat \phi}^{*}+{\hat \pi}_{\phi}\partial^s
\hat \phi ] - (r\leftrightarrow s) \} (\tau ,\vec \sigma ).
\label{b16}
\end{equation}

If we go to the gauge $\chi =T_s-\tau \approx 0$, we can eliminate the variables
$\epsilon_s$, $T_s$, and the $\tau$-evolution (in the Lorentz scalar rest-frame
time) is governed by the Hamiltonian

\begin{eqnarray}
H_R&=&H_{rel}-\vec \lambda (\tau )\cdot {\vec {\cal H}}_p(\tau ),\nonumber \\
&&{}\nonumber \\
H_{rel}&=&{1\over 2}\int d^3\sigma ({\vec \pi}^2
_{\perp}+{\vec B}^2)(\tau ,\vec \sigma )+\nonumber \\
&+&\int d^3\sigma [{\hat \pi}_{\phi^{*}}{\hat \pi}_{\phi}+(\vec \partial +ie
{\vec A}_{\perp}{\hat \phi}^{*}\cdot (\vec \partial -ie{\vec A}_{\perp})\hat 
\phi +m^2{\hat \phi}^{*}\hat \phi ](\tau ,\vec \sigma )-\nonumber \\
&-&{{e^2}\over {8\pi}}\int d^3\sigma_1d^3\sigma_2 {{i({\hat \pi}_{\phi^{*}}{\hat
\phi}^{*}-{\hat \pi}_{\phi}\hat \phi )(\tau ,{\vec \sigma}_1)\, i({\hat \pi}
_{\phi^{*}}{\hat \phi}^{*}-{\hat \pi}_{\phi}\hat \phi )(\tau ,{\vec \sigma)_2)}}
\over {|{\vec \sigma}_1-{\vec \sigma}_2|}}.
\label{b17}
\end{eqnarray}

In the gauge $\vec \lambda (\tau )=0$, the Hamilton equations are

\begin{eqnarray}
\partial_{\tau} \hat \phi (\tau ,\vec \sigma )\, &{\buildrel \circ \over =}\,&
{\hat \pi}_{\phi^{*}}(\tau ,\vec \sigma )+\nonumber \\
&+&{{ie^2}\over {4\pi}}\hat \phi (\tau ,\vec \sigma ) \int d^3\bar \sigma
{{i({\hat \pi}_{\phi^{*}}{\hat \phi}^{*}-{\hat \pi}_{\phi}\hat \phi )(\tau ,
{\vec {\bar \sigma}})}\over {|\vec \sigma -{\vec {\bar \sigma}}|}},
\nonumber \\
\partial_{\tau}{\hat \pi}_{\phi^{*}}(\tau ,\vec \sigma )\, &{\buildrel \circ
\over =}\,& [(\vec \partial -ie{\vec A}_{\perp}(\tau ,\vec \sigma ))^2-m^2]
\hat \phi (\tau ,\vec \sigma )+\nonumber \\
&+&{{ie^2}\over {4\pi}}{\hat \pi}_{\phi^{*}}(\tau ,\vec \sigma )\int d^3\bar 
\sigma {{i({\hat \pi}_{\phi^{*}}{\hat \phi}^{*}-{\hat \pi}_{\phi}\hat \phi )
(\tau ,{\vec {\bar \sigma}})}\over {|\vec \sigma -{\vec {\bar \sigma}}|}},
\nonumber \\
&&{}\nonumber \\
\partial_{\tau}A^r_{\perp}(\tau ,\vec \sigma )\, &{\buildrel \circ \over =}\,&
-\pi^r_{\perp}(\tau ,\vec \sigma ),\nonumber \\
\partial_{\tau}\pi^r_{\perp}(\tau ,\vec \sigma )\, &{\buildrel \circ \over =}\,&
\triangle A^r_{\perp}(\tau ,\vec \sigma )+\nonumber \\
&+&ie P^{rs}_{\perp}(\vec \sigma ) [{\hat \phi}^{*}(\partial^s-ieA^s_{\perp})
\hat \phi -\hat \phi (\partial^s+ieA^s_{\perp}){\hat \phi}^{*}](\tau ,\vec 
\sigma ).
\label{b18}
\end{eqnarray}

\noindent The equations for ${\hat \phi}^{*}$ and $\pi_{\phi}$ are the complex 
conjugate of those for $\hat \phi$ and for ${\hat \pi}_{\phi^{*}}$.

By using the results of Ref.\cite{lv1}, we have the following inversion formula

\begin{eqnarray}
{\hat \pi}_{\phi^{*}}\, &{\buildrel \circ \over =}\,& \partial_{\tau} \hat
\phi +ie^2 \hat \phi {1\over {\triangle}} i({\hat \pi}_{\phi^{*}}{\hat \phi}^{*}
-{\hat \pi}_{\phi}\hat \phi )=\nonumber \\
&=&\partial_{\tau} \hat \phi +ie^2 \hat \phi {1\over {\triangle +2e^2{\hat 
\phi}^{*}\hat \phi}} i({\hat \phi}^{*} \partial_{\tau}\hat \phi -\hat \phi 
\partial_{\tau} {\hat \phi}^{*}),
\label{b19}
\end{eqnarray}

\noindent since we have $i({\hat \phi}^{*}\partial_{\tau}\hat \phi -\hat \phi
\partial_{\tau}{\hat \phi}^{*})=[1+2e^2{\hat \phi}^{*}\hat \phi {1\over 
{\triangle}}] i({\hat \pi}_{\phi^{*}}{\hat \phi}^{*}-{\hat \pi}_{\phi}\hat 
\phi )$ and where use has been done of the operator identity ${1\over A}{1\over
{1+B{1\over A}}}={1\over A}[1-B{1\over A}+B{1\over A}B{1\over A}-...]={1\over
{A+B}}$ (valid for B a small perturbation of A) for $A=\triangle$ and $B=2e^2
{\hat \phi}^{*}\hat \phi$.

Using this formula, we get the following second order equations of motion

\begin{eqnarray}
&&\{ [\, \partial_{\tau}+ie^2{1\over {\triangle +2e^2{\hat \phi}^{*}\hat \phi}}
\, i({\hat \phi}^{*} \partial_{\tau}\hat \phi -\hat \phi \partial_{\tau}{\hat
\phi}^{*})]^2-(\vec \partial -ie{\vec A}_{\perp})^2+m^2\, \} \hat \phi 
{\buildrel \circ \over =}\, 0,\nonumber \\
&&{}\nonumber \\
&&[\partial_{\tau}^2+\triangle ] A^r_{\perp}\, {\buildrel \circ \over =}\,
ie P^{rs}_{\perp}(\vec \sigma ) [{\hat \phi}^{*}(\partial^s -ieA^s_{\perp})\hat
\phi -\hat \phi (\partial^s+ieA^s_{\perp}){\hat \phi}^{*}].
\label{b20}
\end{eqnarray}

We see that the non-local velocity-dependent self-energy is formally playing the
role of a scalar potential.

The previous results can be reformulated in the two-component Feshbach-Villars
formalism for the Klein-Gordon field \cite{fv} [see also Ref.\cite{cs,gross}].
If we put ($\tau_i$ are the Pauli matrices)

\begin{eqnarray}
\hat \phi &=&{1\over {\sqrt{2}}}[\varphi +\chi ],\nonumber \\
{i\over m}{\hat \pi}_{\phi^{*}}&=&{1\over {\sqrt{2}}}[\varphi -\chi ],
\nonumber \\
&&{}\nonumber \\
&&\varphi = {1\over {\sqrt{2}}} [\hat \phi +{i\over m}{\hat \pi}_{\phi^{*}}],
\nonumber \\
&&\chi = {1\over {\sqrt{2}}} [\hat \phi -{i\over m}{\hat \pi}_{\phi^{*}}],
\nonumber \\
&&{}\nonumber \\
\Phi &=&\left( \begin{array}{c} \varphi \\ \chi  \end{array} \right) ,
\label{b21}
\end{eqnarray}

\noindent the Hamilton equations for the Klein-Gordon field become

\begin{eqnarray}
&&i\partial_{\tau}\varphi \,{\buildrel \circ \over =}\, {1\over {2m}}{(-i\vec 
\partial -e{\vec A}_{\perp})}^2(\varphi +\chi )+(m+K)\varphi ,\nonumber \\
&&i\partial_{\tau}\chi =-{1\over {2m}}{(-i\vec \partial -e{\vec A}_{\perp})}^2
(\varphi +\chi )+(-m+K)\chi ,\nonumber \\
&&{}\nonumber \\
&&K(\tau ,\vec \sigma )=-{{me^2}\over {4\pi}} \int d^3\sigma_1 {{(\varphi^{*}
\varphi -\chi^{*}\chi )(\tau ,{\vec \sigma}_1)}\over {|\vec \sigma -{\vec 
\sigma}_1|}}=\nonumber \\
&&=-{{me^2}\over {4\pi}}\int d^3\sigma_1 {{(\Phi^{*}\tau_3\Phi )(\tau ,{\vec
\sigma}_1)}\over {|\vec \sigma -{\vec \sigma}_1|}}.
\label{b22}
\end{eqnarray}

\noindent In the $2\times 2$ matrix formalism we have 

\begin{eqnarray}
i\partial_{\tau} \Phi &=& [{1\over {2m}}{(-i\vec \partial -e{\vec A}_{\perp})}^2
\, (\tau_3+i\tau_2)+m\tau_3+K \openone ]\Phi =\nonumber \\
&=&H \Phi .
\label{b23}
\end{eqnarray}

Since $\rho ={i\over m}\Phi^{*}\tau_3\Phi={i\over m}(\varphi^{*}\varphi -\chi
^{*}\chi )={i\over m}({\hat \pi}_{\phi^{*}}{\hat \phi}^{*}-{\hat \pi}_{\phi}
\hat \phi )$ is the density of the conserved charge e/m (see the Gauss law),
the normalization of $\Phi$ can be taken $\int d^3\sigma (\Phi^{*}\tau_3\Phi )
(\tau ,\vec \sigma )={e\over m}$.

As shown in Ref.\cite{fv}, when we put ${\vec A}_{\perp}=K=0$, the free Klein-
Gordon field has the Hamiltonian  $H_o={ {{\vec p}^2}\over {2m}}(\tau_3+
i\tau_2)+m\tau_3$ in the momentum representation and this Hamiltonian can be 
diagonalized ($p^{\tau}=+\sqrt{m^2+{\vec p}^2}$)

\begin{eqnarray}
H_{o,U}&=&U^{-1}(\vec p)H_oU(\vec p)=p^{\tau}\tau_3=\left( \begin{array}{cc}
\sqrt{m^2+{\vec p}^2}&0\\ 0&-\sqrt{m^2+{\vec p}^2} \end{array} \right) ,
\nonumber \\
&&\Phi_U(\tau ,\vec p)=U^{-1}(\vec p)\Phi (\tau ,\vec p),\nonumber \\
&&i\partial_{\tau}\Phi_U=H_{o,U}\Phi_u,\nonumber \\
&&{}\nonumber \\
&&U(\vec p)={1\over {2\sqrt{mp^o}}}[(m+p^o) 1-(m-p^o)\tau_1],\nonumber \\
&&U^{-1}(\vec p)={1\over {2\sqrt{mp^o}}}[(m+p^o) 1+(m-p^o)\tau_1].
\label{b24}
\end{eqnarray}

Like in the case of the Foldy-Wouthuysen transformation for particles of spin
1/2, also in the spin 0 case the exact diagonalization of the Hamiltonian
cannot be achieved in presence of an arbitrary external electromagnetic field
\cite{fv}.                                   

Now, Eq.(\ref{b23}) has the following form after Fourier transform

\begin{eqnarray}
i\partial_{\tau} \Phi (\tau ,\vec p)&=& \tilde H \Phi (\tau ,\vec p),
\nonumber \\
\tilde H&=&{1\over {2m}} [\vec p-e\int d^3k {\vec A}_{\perp}(\tau ,\vec k)e^{-
\vec k\cdot \vec \partial}\, ]^2(\tau_3+i\tau_2)+m\tau_3+\nonumber \\
&+&\int d^3k K(\tau ,\vec k) e^{-\vec k\cdot \vec \partial}\, \openone .
\label{b25}
\end{eqnarray}

If we put $\Phi(\tau ,\vec p)=U(\vec p) \Phi_U(\tau ,\vec p)$ with the same
$U(\vec p)$ of the free case, we get [see Ref.\cite{fv}]

\begin{eqnarray}
i\partial_{\tau} \Phi_U(\tau ,\vec p)&=& \sqrt{m^2+{\vec p}^2} \tau_3 \Phi
_U(\tau ,\vec p)+\nonumber \\
&+&\int d^3k K(\tau ,\vec p-\vec k){{(\sqrt{m^2+{\vec p}^2}+\sqrt{m^2+{\vec 
k}^2})\openone +(\sqrt{m^2+{\vec p}^2}-\sqrt{m^2+{\vec k}^2})\tau_1}\over
{2\sqrt{ \sqrt{m^2+{\vec p}^2} \sqrt{m^2+{\vec k}^2} } }} \nonumber \\
&&\openone \Phi_U(\tau ,\vec k)+
\nonumber \\
&+&\int d^3k {m\over {2\sqrt{ \sqrt{m^2+{\vec p}^2} \sqrt{m^2+{\vec k}^2} }}}
[-{e\over m}\vec k\cdot {\vec A}_{\perp}(\tau ,\vec p-\vec k)+{{e^2}\over {2m}}
({\vec A}^2_{\perp})(\tau ,\vec p-\vec k)] \nonumber \\
&&(\openone +\tau_1) \Phi_U(\tau 
,\vec k), 
\label{b26}
\end{eqnarray}

\noindent where $({\vec A}^2_{\perp})(\tau ,\vec p)$ means the Fourier 
transform of ${\vec A}^2_{\perp}(\tau ,\vec \sigma )$.

In ref.\cite{fv}, it is shown that this Hamiltonian cannot be diagonalized, 
because the separation of positive and negative energies is inhibited by effects
which (in a second quantized formalism) can be asribed to the vacuum 
polarization, namely to the pair production. This (i.e. the nonseparability
of positive and negative energies) is also the source of the 
zitterbewegung effects for localized Klein-Gordon wave packets as discussed in
Ref.\cite{fv}.

Let us come back to the constraint (\ref{b15}) giving the invariant mass of
the full system. With an integration by parts it can be rewritten as

\begin{eqnarray}
\epsilon_s&-&{1\over 2}\int d^3\sigma ({\vec \pi}_{\perp}^2+{\vec B}^2(\tau ,
\vec \sigma ) -\nonumber \\
&-&\int d^3\sigma \Phi^{*}(\tau ,\vec \sigma )\tau_3 [{1\over 2}(-i\vec
\partial -e{\vec A}_{\perp}(\tau ,\vec \sigma ))^2 (\tau_3+i\tau_2)+m^2\tau_3]
\Phi (\tau ,\vec \sigma )+\nonumber \\
&+&\int d^3\sigma \Phi^{*}(\tau ,\vec \sigma )\tau_3[{{e^2m^2}\over {8\pi}}
\int d^3\sigma_1 {{(\Phi^{*}\tau_3\Phi )(\tau ,{\vec \sigma}_1)}\over
{|\vec \sigma -{\vec \sigma}_1|}} \openone ] \Phi (\tau ,\vec \sigma )\approx 0.
\label{b27}
\end{eqnarray}

If we suppose that $\Phi (\tau ,\vec \sigma )$ is normalized to $\int d^3\sigma
\Phi^{*}(\tau ,\vec \sigma )\tau_3 \Phi (\tau ,\vec \sigma )=1/m$ [this is a
charge normalization compatible with the nonlinear equations of motion, because
the electric charge is conserved], we can rewrite the previous formula as

\begin{eqnarray}
&&\int d^3\sigma \Phi^{*}(\tau ,\vec \sigma )\tau_3 \{ \, [
\epsilon_s-{1\over 2}\int d^3\sigma ({\vec \pi}_{\perp}^2+{\vec B}^2(\tau ,
\vec \sigma ) +\nonumber \\
&+&{{e^2m}\over {8\pi}}
\int d^3\sigma_1 {{(\Phi^{*}\tau_3\Phi )(\tau ,{\vec \sigma}_1)}\over
{|\vec \sigma -{\vec \sigma}_1|}} ] \openone -\nonumber \\
&-&[{1\over {2m}}(-i\vec
\partial -e{\vec A}_{\perp}(\tau ,\vec \sigma ))^2 (\tau_3+i\tau_2)+m\tau_3]
\} \Phi (\tau ,\vec \sigma ) \approx 0.
\label{b28}
\end{eqnarray}

If we assume that the nonlinear equations for the reduced Klein-Gordon field
have solutions of the form $\Phi (\tau ,\vec \sigma )=\Phi (\tau ,\vec p)
e^{i\vec p\cdot \vec \sigma}+\Phi_1(\tau ,\vec \sigma )$ with $\Phi_1$
negligible, namely that the global form of the nonlinear wave admits a sensible
eikonal approximation, then, neglecting $\Phi_1$, we get approximately

\begin{eqnarray}
&&\int d^3\sigma \Phi^{*}(\tau ,\vec \sigma )\tau_3 \{ \, [
\epsilon_s-{1\over 2}\int d^3\sigma ({\vec \pi}_{\perp}^2+{\vec B}^2(\tau ,
\vec \sigma ) +\nonumber \\
&+&{{e^2m}\over {8\pi}}
\int d^3\sigma_1 {{(\Phi^{*}\tau_3\Phi )(\tau ,{\vec \sigma}_1)}\over
{|\vec \sigma -{\vec \sigma}_1|}} ] \openone -\nonumber \\
&-&[{1\over {2m}}(\vec p 
-e{\vec A}_{\perp}(\tau ,\vec \sigma ))^2 (\tau_3=i\tau_2)+m\tau_3]
\} \Phi (\tau ,\vec \sigma ) \approx 0.
\label{b29}
\end{eqnarray}

If we now redefine $\Phi_U(\tau ,\vec \sigma )=U^{-1}(\vec p-e{\vec A}_{\perp}
(\tau ,\vec \sigma )) \Phi (\tau ,\vec \sigma )$ with the same U of 
Eq.(\ref{b24}), we get

\begin{eqnarray}
&&\int d^3\sigma \Phi_U^{*}(\tau ,\vec \sigma )
\left( \begin{array}{cc}{\cal H}_{+}(\tau ,\vec \sigma )&0\\ 
0&{\cal H}_{-}(\tau ,\vec \sigma )\end{array} \right) \,
\Phi_U(\tau ,\vec \sigma )\approx 0,\nonumber \\
&&{}\nonumber \\
&&{\cal H}_{\pm}(\tau ,\vec \sigma )=\epsilon_s\mp
\sqrt{m^2+({\vec p}-e{\vec A}_{\perp}(\tau ,\vec \sigma ))^2}+\nonumber \\
&+&{{e^2m}\over {8\pi}} \int d^3\sigma_1
{{(\Phi^{*}\tau_3\Phi )(\tau ,{\vec \sigma}_1)}\over
{|\vec \sigma -{\vec \sigma}_1|}}
-{1\over 2}\int d^3\sigma_1({\vec \pi}^2_{\perp}+{\vec B}^2)(\tau 
,{\vec \sigma}_1).
\label{b30}
\end{eqnarray}

\noindent where ${\cal H}_{\pm}(\tau ,\vec \sigma )\approx 0$ are the 
constraints (\ref{a30}) for N=1
for the invariant mass of charged scalar particles plus the
electromagnetic field given in Ref.\cite{lus1} for the two possible signs
of the energy $\eta =\pm$. The Klein-Gordon self-energy should go in the 
particle limit (eikonal approximation of filed theory) in the Coulomb
self-energy of the classical particle, which is assent in Ref.\cite{lus1}
because it is regularized by assuming that the particle electric charge Q is
pseudoclassically described by Grassmann variables so that $Q^2=0$.
Therefore, the particle description of Ref.\cite{lus1} is valid only when one
disregards the effects induced by vacuum polarization and pair production and 
uses  a strong eikonal approximation neglecting diffractive effects. It would be
interesting to investigate whether there are special electromagnetic fields 
for which Eq.(\ref{b25}) can be diagonalized with a field-dependent matrix
more general of $U(\vec p)$ and whether Eq.(\ref{b30}) can be generalized
with non-minimal coupling terms in these cases.

\vfill\eject

\section{Conclusions.}

In this paper we ended the rest-frame description of scalar charged, either 
particle or field, systems interacting with the electromagnetic field. For
particles we deduced the final reduced  equations of motion and gave
indications of how to attack the problem of getting an equivalent
Fokker-type descrption. Due to the pseudoclassical nature of the
Grassmann-valued electric charges, the causality problems of the 
Abraham-Lorentz-Dirac equation get a solution without damaging the macroscopic
results of radiation theory based on the Larmor equation.

We left the signs $\eta_i$ of the energies of the particles arbitrary. 
Actually, we can put all signs $\eta_i=+1$, with the classical antiparticles 
moving forward in $\tau$ and having opposite electric charge (opposite ratio
charge/mass) with respect to the classical particles\cite{st,fey,cos}.

We showed which should be the starting point for the connection with the
two-body equations of Refs.\cite{crater,saz,crater1,saz1}: is the Darwin
potential at order $1/c^2$ already contained in our reduced theory? Let us add 
that one can introduce further instantaneous
interactions (besides the Coulomb one) in Eqs.(\ref{a30}), at least for N=2, 
in such a way that the constraints remain first class. As shown in Eq.(111)
of Ref.\cite{lus1} one can introduce any additive potential depending on
$|{\vec \eta}_i(\tau )-{\vec \eta}_j(\tau )|$ in the first os Eqs.(\ref{a30})
and the constraints  remain first class, because the other constraints in
Eq.(\ref{a30}) are interaction independent and contain only ${\check {\vec 
\kappa}}_{+}(\tau )=\sum_{i=1}^N{\check {\vec \kappa}}_i(\tau )$; instead aa 
interaction depending also on the transverse electromagnetic field should 
commute with $\int d^3\sigma ({\check {\vec \pi}}_{\perp}\times {\check {\vec 
B}})(\tau ,\vec \sigma )$. Once one has the first class constraints on the
Wigner hyperplane, one can try to go backward and deduce the constraints on
general spacelike hypersurfaces. This could  simulate more general
interactions maybe coming from a partial diagonalization of
Feshbach-Villars' Eqs.(\ref{b26}) with a consistent truncation of the
off-diagonal terms connected with pair production. Moreover,
one should study the separation of 
positive and negative particle energies in the pairs of first class 
constraints of Refs.\cite{crater,saz}
 to see whether it is possible to get a $4\times 4$ formalism
generalizing Eq.(\ref{b30}) in this two-particle sector. One should obtain
4 first class constraints for the 4 branches of the mass spectrum for
$m_1\not= m_2$ [for $m_1=m_2$ the situation is more complex \cite{lus1}] to
be compared with the results of Ref.\cite{lus1}, remembering that in general
the theory of interacting particles on spacelike hypersurfaces is not
completely equivalent to the one with mass-shell constraints [non-timelike
branches of the mass spectrum are eliminated by construction on the 
hypersurface].

We also gave the rest-frame formulation of scalar electrodynamics and we
showed that the approximation with scalar charged particles emerges in an
eikonal approximation after a Feshbach-Villars reformulation.

Let us remark that in the electromagnetic case all the dressing with Coulomb
clouds [of the scalar particles and of charged Klein-Gordon fields in this
paper and of Grassmann-valued Dirac fields in Ref.\cite{lusa}] are done with the
Dirac phase $\eta_{em}=-{1\over {\triangle}} \vec \partial \cdot \vec A$ 
\cite{di}. The same phase is used in Ref.\cite{lav} to dress fermions in QED.
Also in Ref.\cite{john} the solution of the quantum Gauss law constraint on
Schroedinger functional $\Psi [A]$ in the case of two static particles of
opposite charges, is able to reproduce the Coulomb potential and the Coulomb
self-energy with the same mechanism as in Eq.(\ref{b12}) only if $\Psi [A]=
e^{i\eta_{em}} \Phi [A]$ with $\Phi [A]$ gauge invariant and not with 
$\Psi^{'}[A]=e^{i\int_{x_o}^{x_1} d\vec x\cdot \vec A(x^o,\vec x)} \Phi^{'}[A]$
with a phase factor resembling the Wilson loop operator [one has $\Phi [A] 
=e^{i\int_{x_o}^{x_1}d\vec x\cdot {\vec A}_{\perp}(x^o,\vec x)} \Phi^{'}[A]$,
namely the Wilson line operator has been broken in the gauge part plus the
gauge invariant part using Eq.(\ref{b11})].

The next step would be the elimination of the 3 constraints ${\vec {\cal H}}_p
(\tau )\approx 0$ defining the intrinsic rest frame. 
This requires the introduction of 3 gauge-fixings
identifying the Wigner 3-vector describing the intrinsic 3-center of mass
on the Wigner hyperplane. However, till now these gauge-fixings are known only
in the case of an isolated system containing only particles. When the center of
mass canonical decomposition of linear classical field theories will be
available (see Ref.\cite{lm} for the Klein-Gordon field), its reformulation
on spacelike hypersurfaces will allow the determination of these gauge-fixings
also when fields are present and a Hamiltonian description with only
Wigner-covariant relative variables with an explicit  control on the
action-reaction balance between fields and particles or between two types of
fields.

Finally, one has to start the quantization program of relativistic scalar
charged particles plus the electromagnetic field in the rest-frame Coulomb
gauge based on the Hamiltonian $H_{rel}$ of Eq.(\ref{f12}). On the particle
side, the complication is the quantization of the square roots associated
with the relativistic kinetic energy terms. On the field side, the obstacle
is the absence (notwithstanding the absence of no-go theorems) of a complete
regularization and renormalization procedure of electrodynamics in the
Coulomb gauge: see Refs.\cite{cou,lav} for the existing results for QED.
However, as shown in Refs.\cite{lus1,lusa}, the rest-frame instant form of
dynamics automatically gives a physical ultraviolet cutoff: it is the
M$\o$ller radius $\rho =\sqrt{-W^2}c/P^2=|\vec S|c/\sqrt{P^2}$ ($W^2=-P^2{\vec 
S}^2$ is the Pauli-Lubanski Casimir), namely the classical intrinsic radius of 
the worldtube, around the covariant noncanonical Fokker-Price center of
inertia, inside which the noncovariance of the canonical center of mass ${\tilde
x}^{\mu}$ is concentrated. At the quantum level $\rho$ becomes the Compton 
wavelength of the isolated system multiplied its spin eigenvalue $\sqrt{s(s+1)}$
, $\rho \mapsto \hat \rho = \sqrt{s(s+1)} \hbar /M$ with $M=\sqrt{P^2}$.

\vfill\eject


\begin{references}

\bibitem{lus1}L.Lusanna, Int.J.Mod.Phys. {\bf A12}, 645 (1997).
\bibitem{dira2}P.A.M.Dirac, {\it Rev.Mod.Phys.} {\bf 21} (1949) 392.
\bibitem{dira1}P.A.M.Dirac, "Lectures on 
     Quantum Mechanics", Belfer Graduate School of Science, Monographs Series 
     (Yeshiva University, New York, N.Y., 1964).
\bibitem{kuchar}K.Kuchar, J.Math.Phys. {\bf 17}, 777, 792, 801 (1976); {\bf 18},
1589 (1977).
\bibitem{fv}H.Feshbach and F.Villars, Rev.Mod.Phys. {\bf 30}, 24 (1958).
\bibitem{lv1}L.Lusanna and P.Valtancoli, ``Dirac's Observables for the Higgs
model: I) the Abelian Case", to appear in Int.J.Mod.Phys. A(HEP-TH/9606078).
\bibitem{casalb}R.Casalbuoni, Nuovo Cimento {\bf A33}, 115, 389 (1976). F.A.
Berezin and M.S.Marinov, Ann.Phys.(N.Y.) {\bf 104}, 336 (1977). A.Barducci, 
R.Casalbuoni and L.Lusanna, Nuovo Cimento {\bf A35}, 377 (1976); Nuovo Cimento
Lett. {\bf 19}, 581 (1977); Nucl.Phys. {\bf B124}, 93 (1977).
\bibitem{casala}A.Barducci, R.Casalbuoni and L.Lusanna, Nucl.Phys.
{\bf B180} [FS2], 141 (1981).
\bibitem{lusa}L.Lusanna, Int.J.Mod.Phys. {\bf A10}, 3531 and 3675 (1995).
\bibitem{itz}C.Itzykson and J.B.Zuber, Quantum Field Theory (McGraw Hill,
New York, 1985).
\bibitem{f1}J.L.Anderson and S.Schiminovich, J.Math.Phys. {\bf 8}, 255 (1967).
\bibitem{f4}R.Marnelius, Phys.Rev. {\bf D10}, 2335 (1974).
\bibitem{g1}A.N.Gordeyev, J.Phys. {\bf A8}, 1048 (1975).
\bibitem{g2}A.N.Gordeyev, Theor.Math.Phys. {\bf 36}, 593 (1978).
\bibitem{g3}Ch.G.van Weert, J.Phys. {\bf A9}, 477 (1976).
\bibitem{f2}R.P.Gaida, Yu.B.Klyuchkovskii and V.I.Tretyak, Theor.Math.Phys. 
{\bf 44}, 687 (1980); {\bf 45}, 963 (1980).
\bibitem{f3}J.Llosa and J.Vives, J.Math.Phys. {\bf 35}, 2856 (1994).
X.Ja\'en, R.J\'auregui, J.Llosa and A.Molina, Phys.Rev. {\bf D36},2385 (1987).
X.J\'aen, J.Llosa and A.Molina, Phys.Rev. {\bf D34}, 2302 (1986).
\bibitem{dv}Ph.Droz Vincent, Lett.Nuovo Cimento {\bf 1}, 839 (1969); Phys.Scr.
{\bf 2}, 129 (1970); Ann.Inst.H.Poincar\'e {\bf 27}, 407 (1977) and {\bf 32A},
377 (1980); Phys.Rev. {\bf D19}, 702 (1979).
\bibitem{tod}I.T.Todorov, JINR Report No.E2-10175, Dubna, 1976 (unpublished);
Ann.Inst.H.Poincar\'e {\bf 28}, 207 (1978); ``Constrainy Hamiltonian Approach
to Relativistic Point Particle Dynamics I", SISSA preprint, Trieste 1980;
``Constraint Hamiltonian Mechanics of Directly Interacting Relativistic
Particles" in ``Relativistic Action at a Distance: Classical and Quantum 
Aspects", ed.J.Llosa, Lecture Notes Phys. 162 (Springer, Berlin, 1982).
\bibitem{ll}G.Longhi and L.Lusanna, Phys.Rev. {\bf D34}, 3707 (1986).
\bibitem{crater}H.W.Crater and P.Van Alstine, J.Math.Phys. {\bf 23}, 1997 (1982)
; Ann.Phys.(N.Y.) {\bf 148}, 57 (1983); Phys.Rev. {\bf D30}, 2585 (1984), {\bf
D33}, 1037 (1086), {\bf D36}, 3007 (1987) and {\bf D46}, 766 (1992). H.W.Crater
and D.Yang, J.Math.Phys. {\bf 32}, 2374 (1991).
\bibitem{saz}H.Sazdjian, Phys.Rev.Lett. {\bf 156B}, 381 (1985);
Phys.Rev. {\bf D33}, 3401 (1986); J.Math.Phys. {\bf
28}, 2618 (1987); Ann.Phys.(N.Y.) {\bf 191}, 52 (1989);``Connection of 
Constraint Dynamics with the Bethe-Salpeter Equation", talk at the Int.Symposium
``Extended Objects and Bound States", Karuizawa 1992, eds. O.Hara,
S.Ishida and S.Naka (World Scientific, Singapore, 1992). 
\bibitem{crater1}H.W.Crater and P.Van Alstine, Phys.Rev.Lett. {\bf 53}, 1577 
(1984); J.Math.Phys. {\bf 31}, 1998 (1990); Phys.Rev. {\bf D30}, 2585 (1984); 
{\bf D34}, 1932 (1986); {\bf D36}, 3007 (1987); {\bf D37}, 1982 (1988). 
H.W.Crater, R.L.Becker, C.Y.Wong and P.Van Alstine, 
Phys.Rev {\bf D46}, 5117 (1992). H.W.Crater, C.W.Wong and C.Y.Wong, Found.
Phys. {\bf 24}, 297 (1994).
H.Crater and P.Van Alstine, in ``Constraint's Theory and
     Relativistic Dynamics", Firenze 1986, eds. G.Longhi and L.Lusanna
     (World Scientific, Singapore, 1987) and in ``Constraint Theory and
     Quantization Methods", Montepulciano 1993, eds. F.Colomo, L.Lusanna
     and G.Marmo (World Scientific, Singapore, 1994).
\bibitem{saz1}H.Sazdjian, Phys.Rev. {\bf D33}, 3425 and 3435 (1986); 
J.Math.Phys. {\bf
29}, 1620 (1988). M.Bawin, J.Cugnon and H.Sazdjian, Int.J.Mod.Phys. {\bf A9}, 
5711 (1994); {\bf A11},5303 (1996).
J.Mourad and H.Sazdjian,  J.Math.Phys. {\bf 35}, 6379 (1994);
J.Phys. {\bf G21}, 267 (1995). H.Jallouli and H.Sazdjian, Phys.Lett. {\bf B366},
409 (1996).
\bibitem{gst}R.Giachetti and E.Sorace, Nuovo Cimento {\bf 63B} (1981) 666;
     E.Sorace and M.Tarlini, Nuovo Cimento {\bf 71B} (1982) 98.
\bibitem{cs}E.Corinaldesi and F.Strocchi, Relativistic Wave Equations 
(North-Holland, Amsterdam, 1963).
\bibitem{gross}F.Gross, ``Relativistic Quantum Mechanucs and Field Theory"
(Wiley-Interscience, New York, 1993).
\bibitem{st}E.C.G.Stueckelberg, Helv.Phys.Acta {\bf 15},23 (1942).
\bibitem{fey}R.P.Feynman, Phys.Rev. {\bf 74}, 939 (1948); {\bf 76}, 749 (1949).
\bibitem{cos}J.P.Costella, B.H.J.McKellar and A.Rawlinson, ``Classical 
Antiparticles", Melbourne Univ.preprint UM-P-97/19 (HEP-TH/9704210).
\bibitem{di}P.A.M.Dirac, Can.J.Phys. {\bf 33}, 650 (1955).
\bibitem{lav}
M.Lavelle and D.McMullan, Phys.Rep. {\bf C279},1 (1997).
E.Bagan, M.Lavelle, D.McMullan, B.Fiol and N.Roy, ``How do Constituent Quarks
arise in QCD? Perturbation Theory and the Infra-Red, talk at QCD-96, Montpellier
1996.
E.Bagan, M.Lavelle and D.McMullan, Phys.Lett. {\bf B370}, 128 (1996);
``A Class of Physically Motivated Gauges with
an Infra-Red Finite Electron Propagator", preprint UAB-FT-384, PLY-MS-96-01
(HEP-TH/9602083).
\bibitem{john}P.E.Haagensen and K.Johnson, ``On the Wavefunctional for Two Heavy
Color Sources in Yang-Mills Theory", MIT preprints MIT-CTP-2614 (HEP-TH/9702204)
\bibitem{lm}G.Longhi and M.Materassi, in preparation.
\bibitem{cou}K.Johnson, Ann.Phys.(N.Y.) {\bf 10}, 536 (1960).
R.Hagen, Phys.Rev. {\bf 130}, 813 (1963).
D.Heckathorn, Nucl.Phys. {\bf B156}, 328 (1979).
G.S.Adkins, Phys.Rev. {\bf D27}, 1814 (1983); {\bf D34}, 2489 (1986).
P.J.Doust, Ann.Phys.(N.Y.) {\bf 177}, 169 (1987).
P.J.Doust and J.C.Taylor, Phys.Lett. {B197}, 232 (1987).
J.C.Taylor, in ``Physical and Nonstandard Gauges", eds. P.Gaigg, W.Kummer and
M.Schweda, Lecture Notes Phys. n.361 (Springer, Berlin, 1990), p.137.
G.Leibbrandt, ``Non-Covariant Gauges", ch.9 (World Scientific, Singapore, 1994).


\end{references}
\end{document}